\documentclass[a4paper, 11pt]{article}
\usepackage{amsmath,amsthm,amssymb,mathrsfs,mathtools,graphicx,color}
\usepackage[titletoc,title]{appendix}
\usepackage[numbers ,sort&compress]{natbib}
\usepackage{color}
\newcommand{\nc}{\newcommand}
\nc{\rnc}{\renewcommand}
\nc{\nn}{\nonumber}

\nc{\del}{{\partial}}
\rnc{\Im}{{\mathrm{Im}\,}}
\rnc{\Re}{{\mathrm{Re}\,}}

\nc{\bra}{\langle}
\nc{\ket}{\rangle}

\nc{\tcr}{\textcolor{red}}
\nc{\tcb}{\textcolor{blue}}

\DeclareMathOperator{\End}{End}
\DeclareMathOperator{\Hom}{Hom}

\theoremstyle{definition}

\numberwithin{equation}{section}

\begin{document}

\title{Two-level Quantum Walkers on Directed Graphs I: \\
Universal Quantum Computing}

\author{
Ryo Asaka\thanks{E-mail: hello.ryoasaka@gmail.com}, \,
Kazumitsu Sakai\thanks{E-mail: k.sakai@rs.tus.ac.jp} \, and
Ryoko Yahagi \thanks{E-mail: yahagi@rs.tus.ac.jp}
\\\\
\textit{Department of Physics,
Tokyo University of Science,}\\
 \textit{Kagurazaka 1-3, Shinjuku-ku, Tokyo 162-8601, Japan} \\
\\\\
\\
}

\date{December 15, 2021}

\maketitle

\begin{abstract}
In the present paper, the first in a series of two, we propose a model of 
universal quantum computation using a fermionic/bosonic multi-particle 
continuous-time quantum walk with two internal states (e.g., the spin-up 
and down states of an electron).
A dual-rail encoding is adopted to convert information: a 
single-qubit is represented by the presence of a single quantum walker 
in either of the two parallel paths. We develop a roundabout gate 
that moves a walker from one path to the next, either clockwise or 
counterclockwise, depending on its internal state. It
can be realized by a single-particle scattering on a
directed weighted graph with the edge weights $1$ and $\pm i$.
The roundabout gate also allows the spatial information of the quantum walker
to be temporarily encoded in its internal states.
The universal gates are constructed by appropriately combining 
several roundabout gates, 
some unitary gates that act on the internal states and 
two-particle scatterings on straight paths.
Any ancilla qubit is not required 
in our model.
The computation is done by just passing quantum walkers through properly
designed paths. Namely, there is no need for any time-dependent control.
A physical implementation of quantum random access 
memory compatible with the present model will be considered 
in the second paper (arXiv:2204.08709).
\end{abstract}
\section{Introduction}\label{introduction}

Quantum walks were introduced as a quantum version of random walks
and have since been widely studied in various fields of mathematics, 
physics and computer science \cite{konno2002quantum,kempe2003quantum,
ambainis2003quantum,
kendon2007decoherence,konno2008quantum,venegas2008quantum,
venegas2012quantum,wang2013physical,
portugal2013quantum}. The time evolution of quantum 
walks is generated by a reversible unitary process,  
unlike classical random walks, 
which evolve according to a stochastic process.  Namely, the randomness 
stems from a quantum superposition state due to a unitary 
evolution and its collapse into a particular state with a certain 
probability after an observation or measurement.
Due to their remarkable characteristics, most notably fast-spreading 
properties caused by quantum interference, quantum walks have 
been employed as quantum algorithms that significantly reduce the 
computation time for solving practical problems: a search problem 
\cite{moore2002quantum,shenvi2003quantum,carneiro2005entanglement}, 
a hitting time problem \cite{ambainis2003quantum,childs2003exponential,
carneiro2005entanglement,tregenna2003controlling}, 
an element distinctness problem \cite{ambainis2007quantum,richter2007almost},
and a graph isomorphism problem \cite{childs2003exponential,kempe2003quantum,
gerhardt2764random,shiau2003physically,douglas2008classical,gamble2010two,
berry2011two}.

\begin{figure}[t]
\centering
\includegraphics[width=0.75\textwidth]{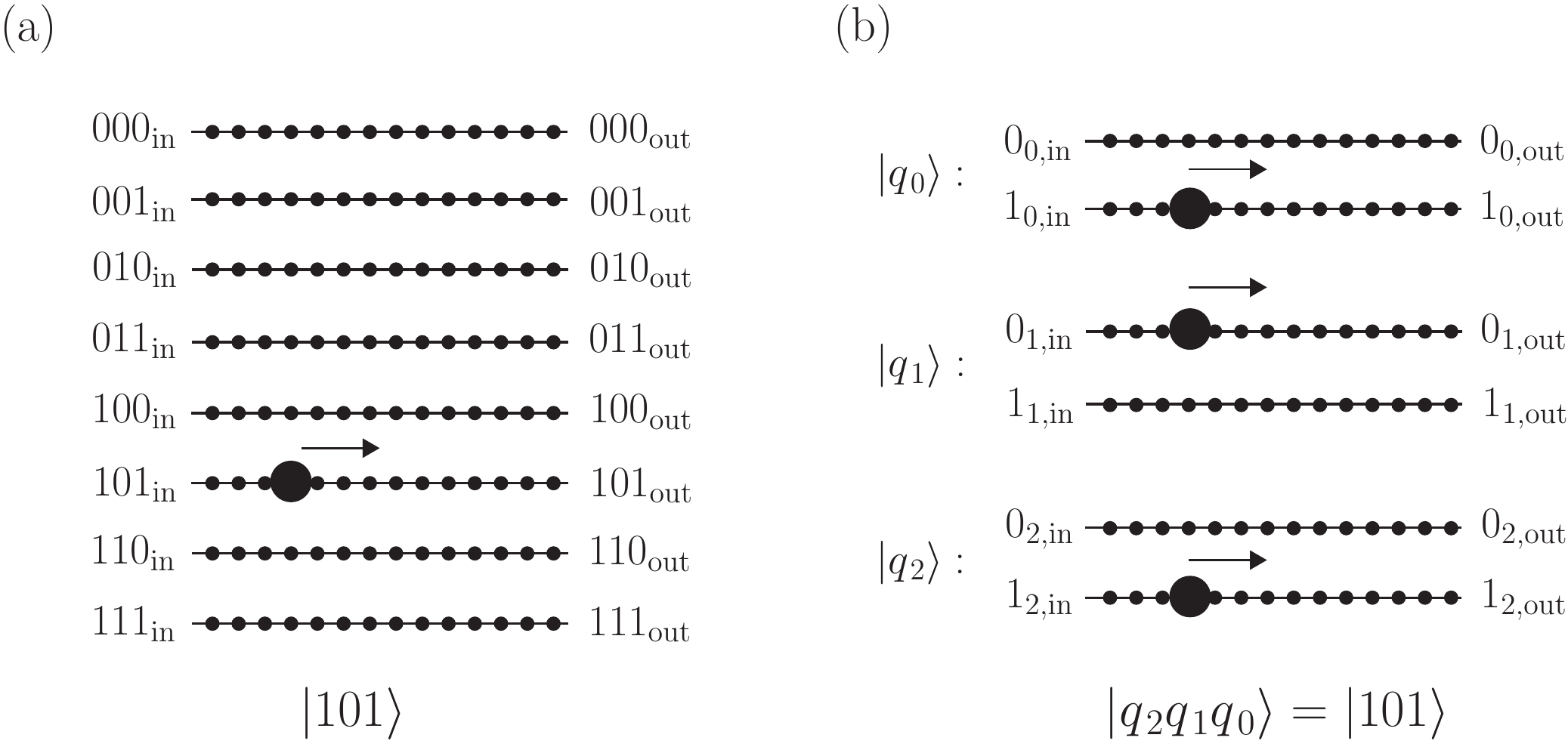}
\caption{
Information spatially encoded by the presence
of a quantum walker. (a): A representation used in a single-particle
architecture \cite{childs2009universal}. $2^n$ paths are required to 
express 
$n$-qubit information.
 (b): A dual-rail encoding used in \cite{childs2013universal}
and the present work. In contrast to (a), only $2n$ paths are
necessary to represent  $n$-qubit information. Instead,  $n$ 
walkers are required. Both (a) and (b)
express the state $|5\ket=|101\ket$. See \eqref{dual1}
and \eqref{dual2}
for more precise definition about the dual-rail encoding.
}
\label{dual-rail}
\end{figure}

On the other hand, quantum walks can also be viewed as an architecture 
of quantum computing. Childs \cite{childs2009universal} has proposed a novel model 
of universal quantum computation using a continuous-time quantum 
walk. A quantum walker evolves continuously in time $t$ 
but discretely in space, 
according to a unitary operator $e^{-i \mathcal{A}(G) t}$ associated with 
the adjacency matrix $\mathcal{A}(G)$ for an unweighted graph $G$.
An $n$ qubit state is represented by a superposition state in  
$2^n$-dimensional Hilbert space, where the basis state is expressed 
as the presence of a single quantum walker on any of the $2^n$ semi-infinite 
paths in $G$ (see Fig.~\ref{dual-rail} (a)). The universal gates 
are implemented by single-particle
scattering processes on a subgraphs $\hat{G}$ connected to $2^n$ semi-infinite
paths. That is, the outcome of the computation corresponds to the final state 
of the following process: (i) a quantum walker starts at one of the semi-infinite 
paths (in superposition) and moves toward $\hat{G}$,  
(ii) it scatters on $\hat{G}$, 
and (iii) it goes out to some of semi-infinite paths.  Shortly after that, 
Lovett et al. extended this idea to a discrete-time version of the quantum walk
\cite{lovett2010universal,underwood2010universal}.

While the universal gates can be constructed via quantum walks, they cannot
be straightforwardly utilized as a practical architecture for a 
quantum computer
because of their lack of scalability: $2^n$ paths are required to 
express $n$ qubits. See Fig.~\ref{dual-rail} (a), for example. 
To overcome the difficulty, introducing multi-particle
continuous-time quantum walks, Childs, Gosset and Webb have designed a 
practical model of a quantum computer \cite{childs2013universal}. 
They used a so-called 
dual-rail encoding where a qubit is denoted by the presence of a single 
quantum walker in either of the two semi-infinite paths (see 
Fig.~\ref{dual-rail} (b)).
Single-qubit gates are implemented in the same way as the single-particle 
universal computation. 
On the other hand, two-qubit gates are realized by a particular combination of
 single-qubit gates and two-particle scattering of a walker representing 
the logical qubit from that for the ancilla qubit 
(see Fig.~\ref{CP} (b) for an implementation of 
the controlled-phase (CP) gate).
Consequently, the graphs required for the architecture are exponentially 
smaller than those for single-particle quantum walks. This remarkable 
progress has been applied to various aspects of quantum computation via 
quantum walks \cite{reitzner2012quantum,bao2015universal,thompson2016time,
piccinini2017gpu,ezawa2019electric,tamura2020quantum,ezawa2020electric,
singh2021universal}.
 
In another direction of application of quantum walks to quantum computation, 
the authors of the present paper have recently proposed a new algorithm for 
quantum random access memory (qRAM) 
\cite{asaka2021quantum}.  A qRAM is a quantum device to access $2^n$ $m$-qubit
data $|x^{(a)}\ket_D\in(\mathbb{C}^2)^{\otimes m}$ ($0\le a \le 2^n-1$)
stored in memory cells at addresses $|a\ket_A\in(\mathbb{C}^2)^{\otimes n}$,
and retrieve the data in superposition:
\begin{equation}
\text{qRAM}:\,\,
\sum_{a}|a \ket_A  |0\ket_D \mapsto \sum_a |a \ket_A |x^{(a)}\ket_D.
\end{equation}
In our algorithm, qRAM is defined on a perfect binary tree with depth $n$, 
where data and addresses are dual-rail encoded  by quantum walkers. 
The walkers moving in qRAM have two internal states, and depending on their states, 
a device equipped at each node on the tree (hereafter referred to as a roundabout gate) 
can properly send the walkers to the designated memory cells placed on the leaves of the tree. 
As a result,  our algorithm requires only $O(n)$ steps and $O(n+m)$ qubit 
resources to access and retrieve the $2^n$ $m$-qubit data.
Characteristically, our algorithm promises to require no time-dependent control: 
qRAM processing is accomplished automatically by simply passing walkers 
through the binary tree. 
However, the physical implementation of qRAM, especially how to realize the 
roundabout gate, has remained a crucial open problem.

The main objectives of the series of two papers are a physical implementation of qRAM
 using  continuous-time multiple quantum walkers with two internal states and 
a design of a universal quantum computer compatible with this qRAM.
The first paper (this one) provides a physical implementation of roundabout
gates and uses them to provide a universal set of quantum gates.
An efficient physical implementation of qRAM will be presented in the second paper 
\cite{asaka2021two}.

In our scheme, a single-qubit is spatially
represented by a dual-rail encoding as in the above model
\cite{childs2013universal}. On the other hand, the quantum walker 
can have two internal states 
$|0\ket_\mathrm{c}$ and $|1\ket_\mathrm{c}$ 
(e.g., the spin-up and  down states
of an electron), 
allowing the physical implementation of a roundabout gate and simplifying the 
architecture. 
On the graph $G$, the quantum walkers evolve according to $e^{-i\mathcal{H}_{G}t}$,
where $\mathcal{H}_G$ is the Hamiltonian associated with the 
adjacency matrix $\mathcal{A}(G)$ and its complex conjugate
$\mathcal{A}(G)^\ast$.
In contrast to the model in \cite{childs2013universal}, single-particle scattering on a
subgraph $\hat{G}$ is used only to implement the roundabout gate 
that moves a walker clockwise or counterclockwise from one path to the next,
depending on the internal state of the walker. A feature of this model is that 
the roundabout gate also
allows the spatial information of the quantum walker to be temporarily 
encoded in its internal states.
Some single-qubit gates are 
implemented by devices  equipped along linear paths, 
acting on the internal states of walkers.
The internal states $|0\ket_\mathrm{c}$ and $|1\ket_\mathrm{c}$ of a walker
may coexist only at the moment when the walker passes through a single-qubit 
gate, and its coherence is completely independent of the other gates.

To implement the roundabout gate, we must  consider a scattering on a directed 
weighted graph $\hat{G}$, which is equivalent to imposing an internal-state-dependent 
phase factor on the Hamiltonian $\mathcal{H}_G$. That is,  
in the subgraph $\hat{G}$, a walker with $|0\ket_\mathrm{c}$ is evolved by 
$e^{-i\mathcal{A}(\hat{G})t}$,
while a walker with $|1\ket_\mathrm{c}$ is evolved by  
$e^{-i\mathcal{A}(\hat{G})^\ast t}$. This difference allows for the 
implementation of the roundabout gate.
A two-qubit gate is simply realized by an appropriate combination 
of roundabout gates, single-qubit gates, and two-particle scatterings
of fermionic or bosonic walkers with the same internal state. 
(For instance, see Fig.~\ref{CP} (a)
for the CP gate.) Notably, the calculation is achieved by 
simply passing quantum walkers through adequately designed paths: there is 
no need for any time-dependent control.
Furthermore, any ancilla qubit required in the model of 
\cite{childs2013universal}
to implement the two-qubit gates is unnecessary for our 
architecture. Consequently, we can simplify the architecture
compared to the model in \cite{childs2013universal}.

The remainder of the paper is organized as follows. In the subsequent
section, we define continuous-time multiple quantum walkers with two
internal states. The roundabout gate, playing a pivotal role in our work,
is considered in Sec.~3. In Sec.~4, the universal gates are physically 
implemented by appropriate combinations  of roundabout gates, single-qubit gates, and
two-particle scatterings. 
Sec.~5 describes how to construct a practical circuit using a set of 
quantum gates developed in Sec.~4.
Sec.~6 is devoted to a summary. Some applications
using the roundabout gate are also discussed.
Technical details about scattering theory required in our paper are
left in Appendix.

\section{Two-level quantum walkers on directed graphs}\label{qwalk}
\subsection{General overview}
%
\begin{figure}[tht]
\centering
\includegraphics[width=0.77\textwidth]{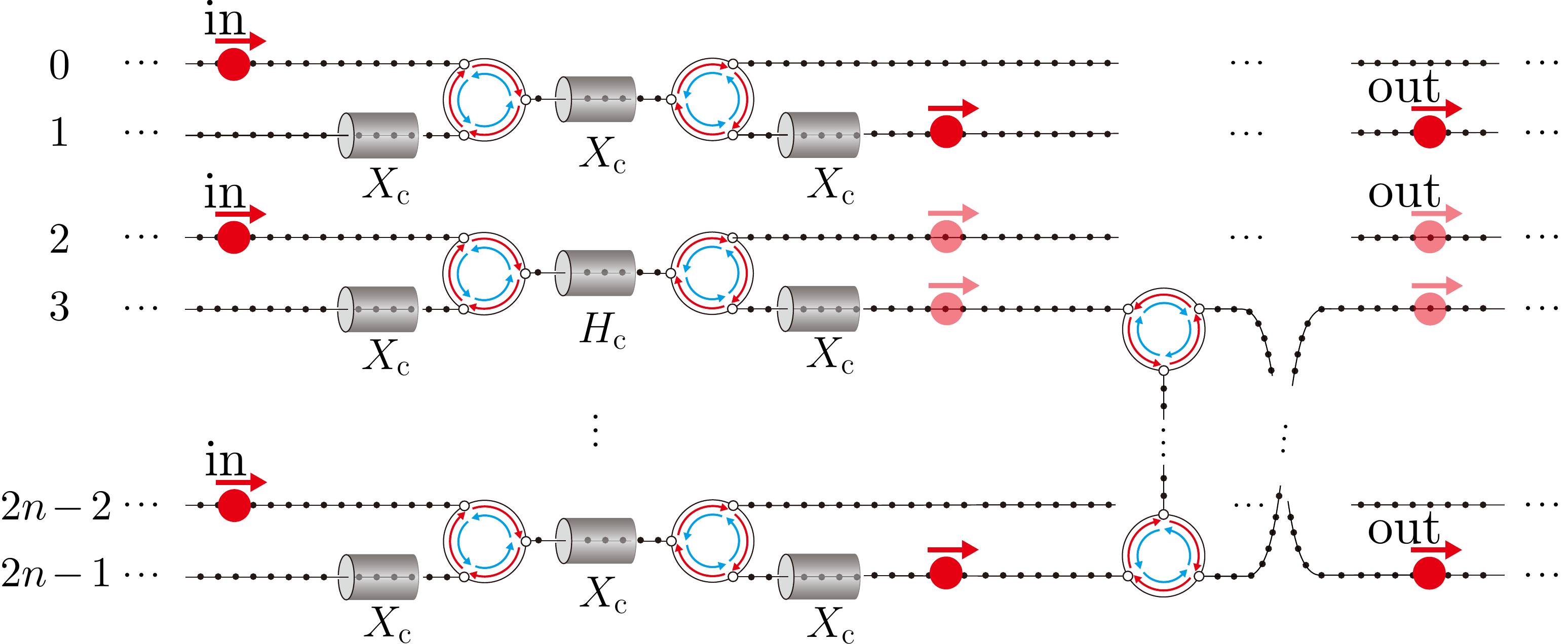}
\caption{An overview of our architecture for quantum computation. 
Initial information is represented by the positions of 
the quantum walkers with internal state  $|0\ket_\mathrm{c}$ (see
eqs. \eqref{dual1} and \eqref{dual2}), 
which are indicated by the red walkers on the left side of the figure.
The walkers (wave packets) have momenta $k\simeq-\pi/2$, and
move to the right along the semi-infinite paths with group 
velocity $v_\mathrm{g}=E'(-\pi/2)=2$, where
$E(k)=2\cos k$ (eq.~\eqref{energy}) is the energy of each walker.
The quantum gates, set up like tunnels, act on the internal states of 
the walkers passing through them. 
Some two red walkers are scattered from each other on a vertical path
connecting  two paths via roundabout gates represented by
\eqref{eq-RA} and \eqref{fig-RA}. After the scattering, only 
the global phase of the two-particle wave function is changed. The position state
of the quantum walkers after passing through the graph 
corresponds to the outcome of the computation. Note that the colors of the
walkers are temporarily mixed only within single-qubit gates and 
are set to red otherwise. The colors do not change 
within the roundabout gates and during the two-particle scatterings.
}
\label{architecture}
\end{figure}
%

First, let us present a general overview of our quantum computing 
architecture via continuous-time quantum walks with two internal states.
As shown in Fig.~\ref{architecture},  a quantum circuit is spatially 
designed on a graph  on which multiple quantum walkers evolve
continuously in time $t$. The dual-rail encoding is employed to
represent a single-qubit state as the presence of
the quantum walker in one of the two parallel paths.
More specifically, a single-qubit state $|q_j\ket$ ($q_j\in\{0,1\}$) is
given by 
\begin{equation}
|q_j\ket=\delta_{q_j,0}|2j\ket_\mathrm{p}+\delta_{q_j,1}|2j+1\ket_\mathrm{p}
\in\mathbb{C}^2,
\label{dual1}
\end{equation}
where $|j\ket_\mathrm{p}$ ($j\in\{0,\cdots,N-1\}$) denotes the state in which 
a quantum walker is located on the $j$th path. Thus an $n$-qubit state
$|q_{n-1}\cdots q_0\ket$ can be expressed by
which of the $2n$ paths the $n$ walkers are located on. Namely,
\begin{equation}
|q_{n-1}\cdots q_0\ket=\bigotimes_{j=0}^{n-1} 
\Bigl(\delta_{q_j,0}|2j\ket_\mathrm{p}+\delta_{q_j,1}|2j+1\ket_\mathrm{p}\Bigr)
\in{(\mathbb{C}^2)}^{\otimes n}.
\label{dual2}
\end{equation}
Note that here and in what follows, we sometimes ignore normalization factors
for simplicity of the notation. In our architecture, the quantum 
walkers have two
internal states $|0\ket_\mathrm{c}\in\mathbb{C}^2$  and  $|1 \ket_\mathrm{c}\in
\mathbb{C}^2$, which 
are normally set to $|0\ket_\mathrm{c}$ except during processing. 
For convenience, the quantum walker whose internal state is $|0\ket_\mathrm{c}$ 
(resp. $|1\ket_\mathrm{c}$) 
is  referred to as the ``red quantum walker" 
(resp. ``blue quantum walker"). 

Roughly, the computation proceeds as follows.
An $n$-qubit input state is prepared by a 
position state of the red quantum walkers (see the left side in 
Fig.~\ref{architecture}). Then, the walkers move right along the paths,
some of which are connected to roundabout gates 
consisting of a subgraph $\hat{G}_\mathrm{R}$.  The walker passing through
the roundabout gate moves from one path to the next in a clockwise 
or counterclockwise direction, 
according to the internal state of the walker. Its unitary operator explicitly
reads
\begin{align}
&U_\mathrm{R}^{(\mathrm{l})}=|0\ket\bra 0|_\mathrm{c} U_{\mathrm{R}}+
  |1\ket\bra 1|_\mathrm{c} U^{\dagger}_{\mathrm{R}},\quad
U_{\mathrm{R}}^{(\mathrm{r})}=U_{\mathrm{R}}^{(\mathrm{l})}{}^{\dagger},\nn \\ 
&U_{\mathrm{R}}=\sum_{k,l=0}^{2}\delta_{l,k+1}|j_l\ket\bra j_k|_\mathrm{p}\quad
(k,l\in\mathbb{Z}/3\mathbb{Z}=\{0,1,2\}),
\label{eq-RA}
\end{align}
where  $U_\mathrm{R}^{(\mathrm{l})}$ moves the red walker (resp. blue walker) 
counterclockwise (clockwise) to the next path,
while  $U_{\mathrm{R}}^{(\mathrm{r})}$ does the opposite to 
$U_{\mathrm{R}}^{(\mathrm{l})}$.
They can be graphically depicted as
\begin{equation}
\includegraphics[width=0.55\textwidth]{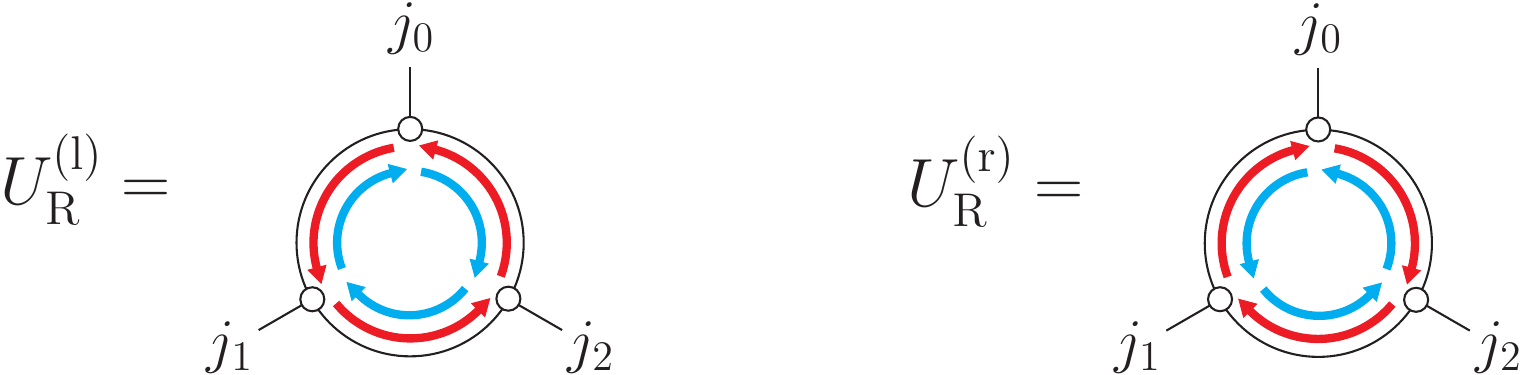}.
\label{fig-RA}
\end{equation}
For instance, the red walker (resp. blue walker) 
entering the $U_\mathrm{R}^{(\mathrm{l})}$ 
(resp. $U_\mathrm{R}^{(\mathrm{r})}$ )
gate from path $j_1$ (resp. $j_2$) exits to
path $j_2$ (resp. $j_0$):
\begin{equation}
\includegraphics[width=0.5\textwidth]{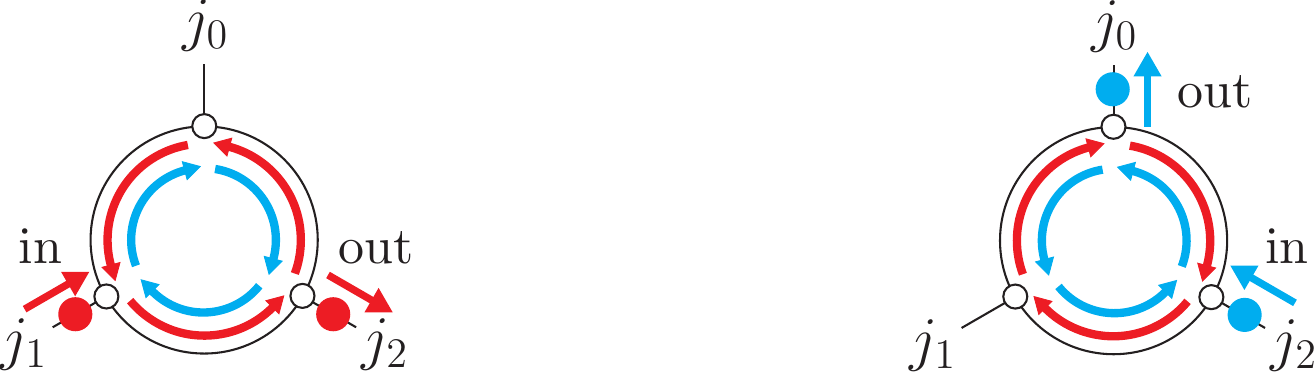}.
\label{fig-RA2}
\end{equation}

The actual implementation of the roundabout gate 
is achieved by a scattering of a single walker with momentum $k=-\pi/2$ 
from the graph $\hat{G}_{\mathrm{R}}$ (see Sec.~\ref{RA} and Appendix).
During the scattering, the internal state of the walker never changes,
but only by some quantum gates set up like tunnels along the path.
For instance, the actions of the Hadamard gate $H_\mathrm{c}$ and 
the Pauli X gate $X_\mathrm{c}$
on the 
state $|0\ket_\mathrm{c} |j\ket_\mathrm{p}$ are respectively written as
\begin{equation}
H_\mathrm{c} |0\ket_\mathrm{c}|j\ket_\mathrm{p}=\frac{1}{\sqrt{2}}\bigl(|0\ket_\mathrm{c}+|1\ket_\mathrm{c}\bigr)|j\ket_\mathrm{p},
\quad
X_\mathrm{c} |0\ket_\mathrm{c}|j\ket_\mathrm{p}=|1\ket_\mathrm{c}|j\ket_\mathrm{p}.
\label{Pauli}
\end{equation} 
They are also graphically shown as
\begin{equation}
\includegraphics[width=0.7\textwidth]{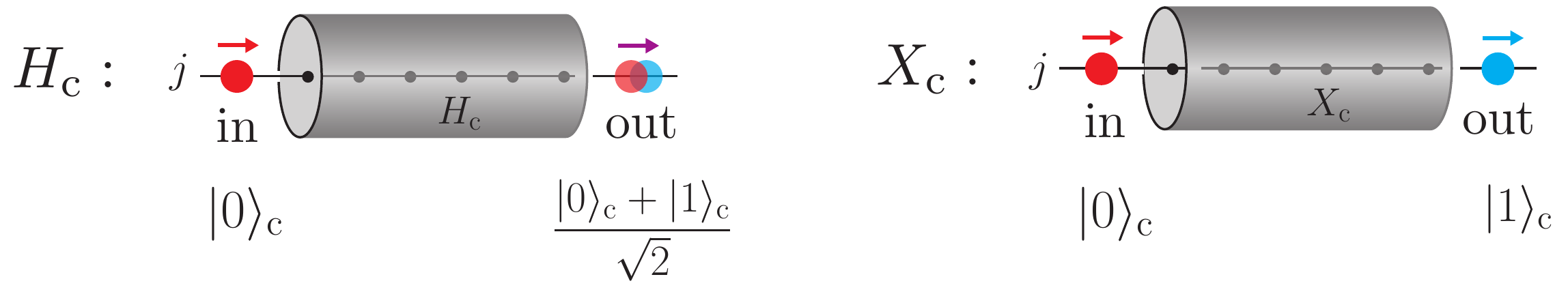}.
\label{fig-gate}
\end{equation}
The roundabout gate also serves as an information encoder and decoder:
spatial information of the quantum walker can be encoded in 
the internal state of the walker. For instance,
\begin{equation}
  \includegraphics[width=0.44\textwidth]{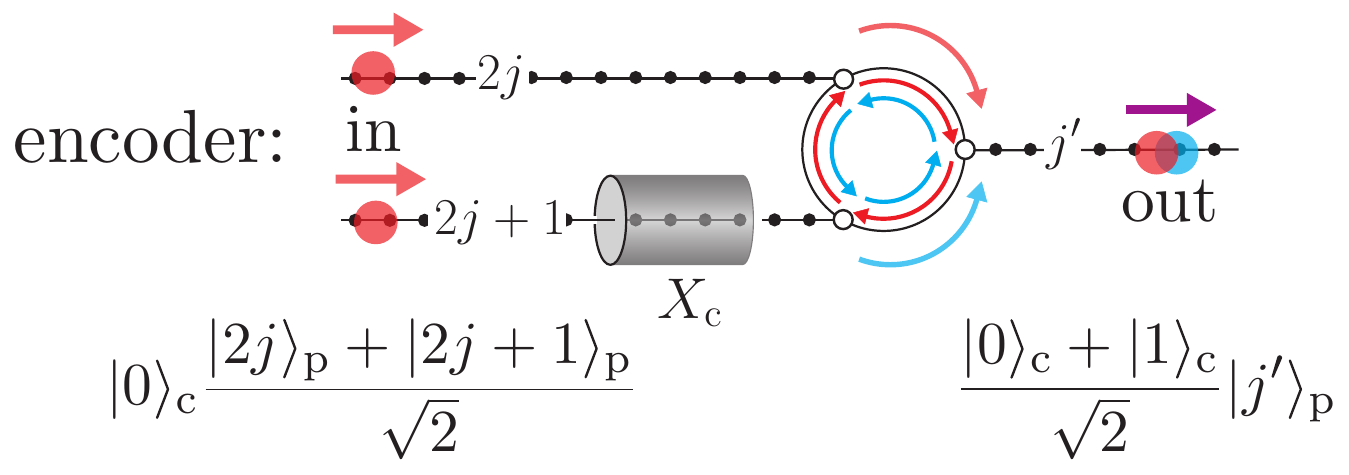},
  \includegraphics[width=0.49\textwidth]{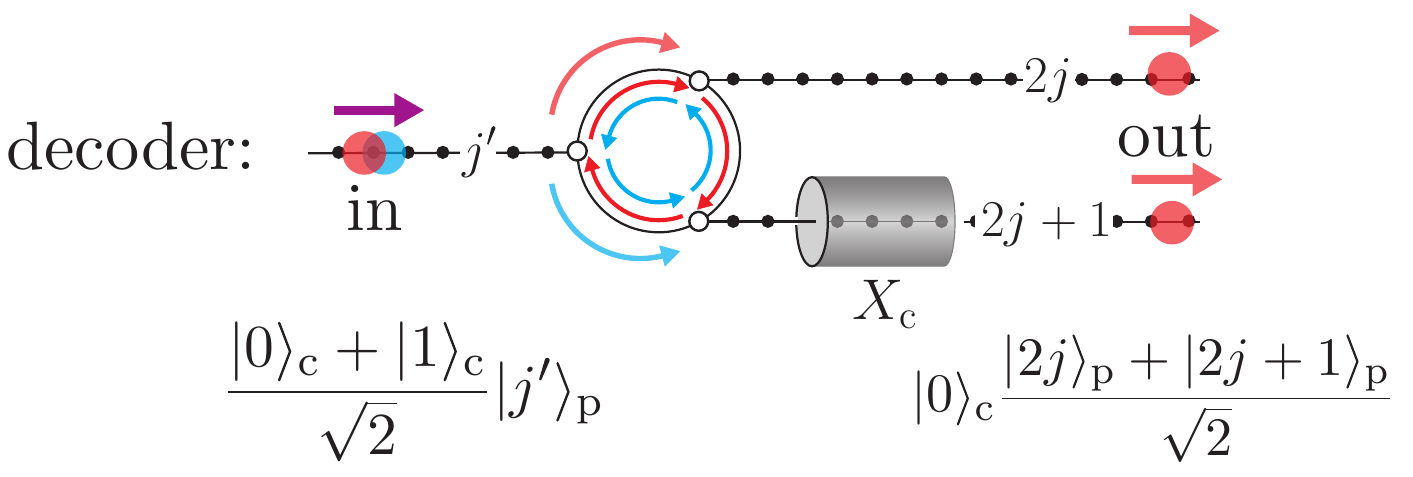}.
  \label{fig-encode/decode}
\end{equation}
Namely, using the encoder and decoder, we can replace any unitary transformation 
of single-qubit information with a transformation of the internal state of the
quantum walker. Thus, in our architecture, the only non-trivial graph required for 
quantum walks is the graph used in the roundabout gate,  simplifying the structure 
of the quantum circuit.

See Sec.~\ref{EG} for more details about the implementation.
Some two walkers with the same internal state may be scattered from each other on  
a straight path (a vertical path in Fig.~\ref{architecture}) connecting
two paths through roundabout gates. Since the total energy and momentum are conserved, 
the individual momenta of the two walkers are  conserved
even after the scattering. As a result, only the global phase
of the two-particle wave function
can be changed after the scattering. See Sec.~\ref{EG} and Appendix
for the calculation of two-particle scattering processes.
The final state obtained after these processes
 (see the right side in Fig.~\ref{architecture}) corresponds to 
the outcome of the computation. Note again that, in our architecture, the colors 
of the walkers are in superposition only within each single-qubit gate and 
are set to red otherwise. In other words, maintaining the coherence of 
the colors is only necessary within individual single-qubit gates.
In actual quantum circuits, the quantum gates must be arranged 
so that two-particle scatterings occur at the 
appropriate positions, which will be discussed in Sec.~\ref{building a circuit}.
\subsection{Quantum walkers on directed graphs}
Now, let us explain the details about the evolution of the 
quantum walkers on directed weighted graphs. As indicated earlier, 
spatial dynamics of the quantum walkers required for the
computation can be essentially decomposed into  motions on
semi-infinite paths, single-particle scatterings on  subgraph
$\hat{G}_{\mathrm{R}}$, and scatterings of  two-walkers
with same internal state. Hence, the procedure developed 
in \cite{childs2013universal,childs2012levinson} is directly applicable 
to describe 
the spatial dynamics of our model.  Here and in Appendix,
we summarize the procedure to make our paper self-contained and 
fix the notation.

\begin{figure}[t]
  \centering
  \includegraphics[width=1\textwidth]{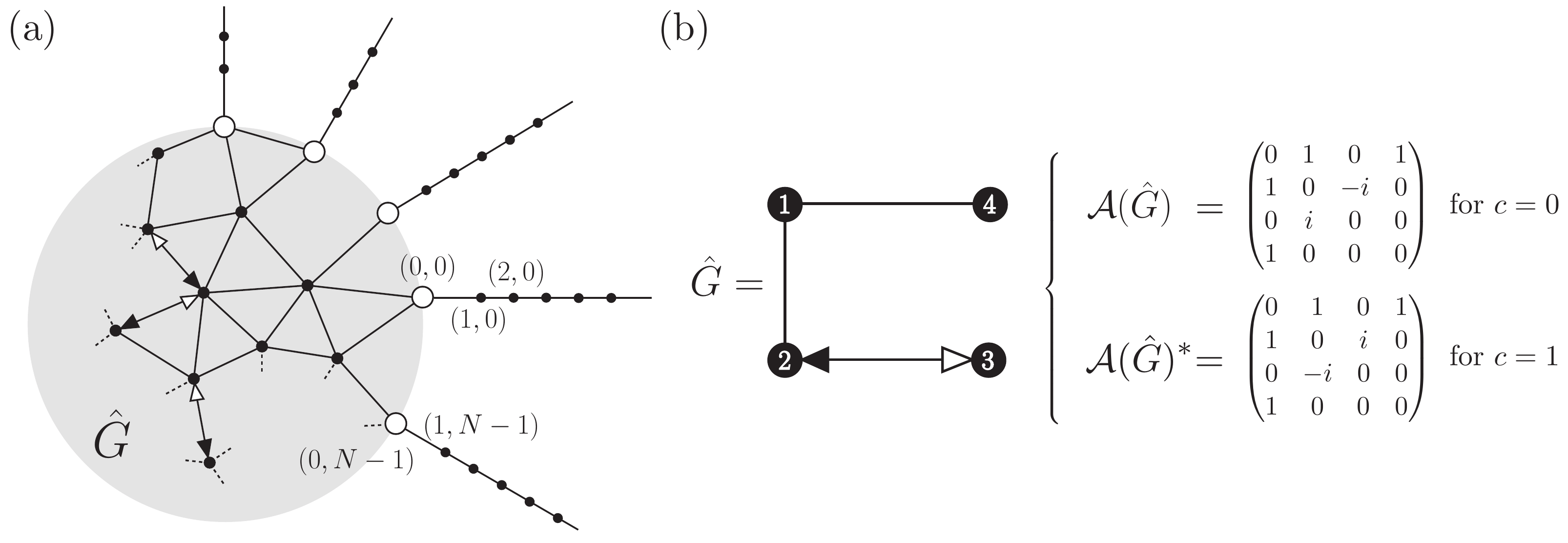}
  \caption{
  (a): A schematic graph $G$ consisting of a subgraph $\hat{G}$ and
  $N$ semi-infinite paths connected to $\hat{G}$ (depicted on the gray
  disk) at the terminals
  denoted by the white circles. The vertices on the $j$th semi-infinite
  path are labeled by $(x,j)$ ($x\in\mathbb{Z}_{\ge 0}$, 
  $j\in\{0,\cdots,N-1\}$).
  (b): An example of the directed subgraph $\hat{G}$ and 
its adjacency matrix $\mathcal{A}(\hat{G})$. As shown in \eqref{kinetic-adjacency},
in the subgraph $\hat{G}$, a single red walker ($c=0$) evolves in time according 
to $e^{-i\mathcal{A}(\hat{G})t}$, 
while a blue walker ($c=1$) evolves according to
 $e^{-i\mathcal{A}(\hat{G})^\ast t}$.}
  \label{graph-G}
  \end{figure}

To formulate the dynamics systematically, we consider the
time evolution of the multiple quantum walkers on a generic 
graph $G=(V,E,w)$ as in \cite{childs2013universal,childs2012levinson}. Here  
$V(G)$ denotes the set of the vertices of $G$, 
$E(G)\subset V\times V$ is the set of the edges. $w:E\to\mathbb{C}$ 
is a function defined by $w(x,y)=w_{xy}$ ($x,y\in V(G)$, $(x,y)\in E(G)$), 
and here is assumed to be 
\begin{equation}
w_{xy}=e^{i\theta_{xy}} \,\, (\theta_{xy}\in\mathbb{R}),
\quad
w_{yx}=w_{xy}^{\ast}=e^{-i\theta_{xy}}, \quad  w_{xx}=0.
\end{equation} 
The weighted adjacency matrix $\mathcal{A}(G)=(a_{xy})$ is defined by 
$a_{xy}:=w_{xy}$. In this work, however, we only use the weights 
$w_{xy}=1$ ($\theta_{xy}=0$) in the semi-infinite paths 
and $w_{xy}=1$ or $w_{xy}=\pm i$ ($\theta_{xy}=\pm\pi/2$) in a subgraph 
$\hat{G}$,
where $(x,y)\in E(G)$. 
As shown in Fig.~\ref{graph-G}~(a), $G$ consists of a subgraph $\hat{G}$
composed by internal vertices, and $N$ semi-infinite 
paths attached to $\hat{G}$ at the terminal vertices of $\hat{G}$ 
(depicted them by white circles in Fig.~\ref{graph-G}~(a)). 
 Let $(x,j)$ ($x\in\mathbb{Z}_{\ge 0}$, $j\in\{0,\cdots,N-1\}$) be the
label of the vertex on the $j$th semi-infinite path, 
located at distance $x$ from $\hat{G}$.  
$T:=\{(0,j)| 0\le j\le N-1\}$  denotes the set of the
terminal vertices of $\hat{G}$.

Let us formulate the dynamics of quantum walkers on the graph $G$, which  is 
governed by the Hamiltonian
\begin{align}
\mathcal{H}_G=\mathcal{K}_G+\mathcal{U}_G.
\label{Hamiltonian}
\end{align}
Here
$\mathcal{K}_G$ is a kinetic term describing non-interacting walkers,
which is defined by
\begin{equation}
\mathcal{K}_G=
\sum_{c=0,1}\sum_{(x,y)\in E(G)} \left(e^{i (-1)^c  \theta_{xy}} 
a_{x,c}^\dagger a_{y,c}+
e^{-i(-1)^c \theta_{xy}} a_{y,c}^{\dagger} a_{x,c}\right),
\label{kinetic term}
\end{equation}
and $\mathcal{U}_G$ denotes  multiple-particle interactions  as explained below
(see \eqref{interaction}). In the summation, we do not distinguish 
between $(x,y)$ and $(y,x)$: the sum is taken over either $(x,y)$ or $(y,x)$.
$a^{\dagger}_{x,c}$ and $a_{x,c}$ are, respectively, the 
creation and annihilation operators of walkers of 
color $c\in\{0,1\}$ on the vertex $x\in V(G)$. The operators satisfy
\begin{equation}
[a_{x,c},a^{\dagger}_{y,c'}]_{\mp}=\delta_{x,y}\delta_{c,c'}, \quad
[a_{x,c},a_{y,c'}]_{\mp}=[a_{x,c}^{\dagger},a_{y,c'}^{\dagger}]_{\mp}=0,
\end{equation}
where $[\cdot,\cdot]_-$ and   $[\cdot,\cdot]_+$ denote the commutator 
for the boson operators
and anticommutator for the fermion operators, respectively. 
The creation operators $a_{x,c}^{\dagger}$ generate the Hilbert
space $H^{(m)}$ for $m$ quantum walkers on $G$: $H^{(m)}$  
is spanned by the basis vectors
\begin{equation}
\left.\left\{
|c_{1}\cdots c_{m}\ket_\mathrm{c}|x_1\cdots x_m\ket:=
a^{\dagger}_{x_m,c_m}\cdots a^{\dagger}_{x_1,c_1}|0\ket\,
\right|c_1,\cdots,c_m\in\{0,1\},\,x_1,\cdots, x_m\in  V(G) 
\right\}, 
\end{equation}
where $|0\ket$ is the no-walker state (vacuum state) defined by 
$a_{x,c}|0\ket=0$
$(c\in \{0,1\}\,\,\,x\in V(G))$. The notation 
$|x,j\ket$ is also used to denote the walker at $(x,j)$, i.e.,
at $x\in\mathbb{Z}_{\ge 0}$ on path $j$ ($j\in\{0,\cdots,N-1\}$).
The time-evolution of the $m$ quantum walkers on the graph $G$ is described
by the action of the unitary operator $e^{-i\mathcal{H}_G t}$
on a vector in the Hilbert space $H^{(m)}$.
Information of the adjacency matrix 
$\mathcal{A}(G)=(e^{i\theta_{xy}})$ ($\theta_{xy}\in \{0,\pm \pi/2\}$, 
$(x,y)\in E(G)$) 
of the graph $G$ is incorporated into the 
kinetic term $\mathcal{K}_G$ \eqref{kinetic term} as
\begin{align}
(\mathcal{K}_G)_{cc',\,xy}&:={}_\mathrm{c}
\bra c| \bra x | \mathcal{K}_G |c'\ket_{\mathrm{c}}|y\ket\nn \\
&=(e^{i\theta_{xy}}\delta_{c,0}\delta_{c',0}+e^{-i\theta_{xy}}
\delta_{c,1}\delta_{c',1})\delta_{(x,y)\in E(G)}\nn \\
&=(\mathcal{A}(G))_{xy}\delta_{c,0}\delta_{c',0}+(\mathcal{A}(G)^\ast)_{xy}
\delta_{c,1}
\delta_{c',1}.
\label{kinetic-adjacency}
\end{align}
Importantly, in the directed weighted graph $\hat{G}$, 
the time-evolution of a walker depends explicitly on its
internal state: the evolution of a red walker (i.e., $c=0$) 
is described by the unitary operator $e^{-i\mathcal{A}(\hat{G})t}$, 
while that of a blue walker (i.e., $c=1$) 
is described by  $e^{-i\mathcal{A}(\hat{G})^\ast t}$.
See also Fig.~\ref{graph-G} (b) for an example of $\hat{G}$.
It is this difference that makes the physical 
implementation of the roundabout gate possible.
Physically, the weight
$w_{xy}=e^{i\theta_{xy}}$ in \eqref{kinetic term} can be interpreted as a 
phase factor 
under a local gauge transformation
$a_{x,c}\mapsto e^{i\vartheta_{x,c}} a_{x,c}$ ($\vartheta_{x,c}\in\mathbb{R}$),
which may be achieved, for example, by applying the Aharonov-Kasher effect
\cite{aharonov1984topological,zvyagin1992aharonov}.

Two-qubit gates are realized by taking into account
two-particle scatterings of fermionic or bosonic walkers 
with the same internal state. To this end,
as in \cite{childs2013universal}, 
we adopt the on-site interaction (Bose-Hubbard model)
for the bosonic case and the nearest-neighbor 
interaction (extended Hubbard model) 
for the fermionic case:
\begin{equation}
\mathcal{U}_G=
\begin{dcases}
\frac{u}{2}\sum_{x\in V(G)}n_x(n_x-1) &\text{ for the bosonic walkers} 
\\
u \sum_{(x,y)\in E(G)}n_x n_y  &\text{ for the fermionic walkers}
\end{dcases},
\label{interaction}
\end{equation}
where $n_x:=\sum_{c=0}^1a_{x,c}^{\dagger}a_{x,c}$ is the number operator.

As an actual implementation, we consider a quantum walker 
as a wave packet $|\Psi,t\ket=e^{-i\mathcal{H}_G t}|\Psi\ket$ 
constructed by a superposition of plane
waves with momentum $k$ close to a specific value of $k_\mathrm{p}$. For instance,
a wave packet on a semi-infinite path (the actual implementation uses a 
sufficiently long path) toward a subgraph $\hat{G}$ 
is given by
\begin{align}
|\Psi,t\ket&=\sum_x\frac{1}{\sqrt{2\pi}}
\int_{k\simeq k_\mathrm{p}} dk f(k) e^{-i k x-i E(k) t}|c\ket_\mathrm{c}|x\ket\nn \\
&\simeq
\sum_x
\frac{e^{-ik_\mathrm{p} x-i E(k_\mathrm{p})t}}{\sqrt{2\pi}}
\int_{k\simeq k_\mathrm{p}} dk 
f(k)e^{-i(k-k_\mathrm{p})(x+E'(k_\mathrm{p})t)}|c\ket_\mathrm{c}|x\ket,
\label{wave-packet}
\end{align}
where $f(k)$ is the Fourier coefficient with a sharp peak at $k\simeq k_\mathrm{p}$, 
and $E(k)=2\cos k$ (eq.~\eqref{energy}) is the energy of the quantum 
walker  on the semi-infinite path.
The wave packet moves with the group velocity $E'(k_\mathrm{p})=-2\sin k_\mathrm{p}$. 
To make our purpose, in the current work, we 
assign $k\simeq k_\mathrm{p}=-\pi/2$ to the momentum of each walker. 

Quantum computation is performed by applying unitary transformations to 
wave packets \eqref{wave-packet}. What is crucial then is how to 
add the overall phase factor to each wave packet (note that simply 
translating the wave packet \eqref{wave-packet} as 
$x\mapsto x+a$ does not yield this), and how to superpose them.
In our architecture, for a 
single quantum walker, this manipulation is essentially accomplished by unitary transformations 
to the internal state of the walker, and when two walkers are involved, as 
in controlled gates, this is accomplished by two-particle scattering.
Single-particle scattering is used only for the implementation of the roundabout 
gate whose primary role is to switch the position of the walker. See Sec.~4 for details. 
This is in contrast to the architecture \cite{childs2013universal}, where all
unitary transformations are performed by single-particle and two-particle scatterings.
In Appendix, we summarize single- and two-particle scatterings  with 
a definite momentum $k$ (i.e., scatterings of plane waves)  which  are 
key to understanding the scattering 
processes of wave packets with momentum close to $k$.

\section{Roundabout gate} \label{RA}
The roundabout gate \eqref{eq-RA} or \eqref{fig-RA} is the most crucial element in 
our architecture.  As schematically depicted in \eqref{fig-RA2}, 
the walker passing through the roundabout gate can move from 
one path to the next. In this section, we implement 
the roundabout gate $U_\mathrm{R}^{(\mathrm{l})}$ with single-particle scattering:
we find a graph $\hat{G}_\mathrm{R}$ such that the $S$-matrix $S(k)$ 
(resp. $\tilde{S}(k)$) 
describing the scattering of the red walker (resp. blue walker)  on $\hat{G}_\mathrm{R}$ 
satisfies $S(k)=U_{\mathrm{R}}$
(resp. $\tilde{S}(k)=U^{\dagger}_{\mathrm{R}}$)
for some specific value of $k$,
where $U_{\mathrm{R}}$ is defined by \eqref{eq-RA}. In fact, the scattering process only 
for the red walker is sufficient to study, because
the $S$-matrix $\tilde{S}(k)$ for the blue walker is nothing but 
the transpose of the $S$-matrix $S(k)$ for the red walker, as 
shown in Appendix (see \eqref{transpose}). Additionally, the 
color of the walkers does not change during the 
single-particle scattering, which can be seen by the Hamiltonian 
\eqref{Hamiltonian}, especially the kinetic term
\eqref{kinetic term}.
A physical implementation of another type of  
roundabout gate $U_{\mathrm{R}}^{(\mathrm{r})}(:=
U_{\mathrm{R}}^{(\mathrm{l})}{}^{\dagger})$ in 
\eqref{eq-RA} can be 
easily accomplished by replacing the subgraph 
$\hat{G}_\mathrm{R}$ with $\hat{G}^{\ast}_\mathrm{R}$
whose adjacency matrix is given by $\mathcal{A}(\hat{G}^{\ast}_\mathrm{R})
=\mathcal{A}(\hat{G}_\mathrm{R})^\ast$.

The $S$-matrix $S(k)$ of the red walker characterizing the scattering 
state is expressed in terms of the elements of the adjacency matrix 
$\mathcal{A}(\hat{G}_\mathrm{R})$, as shown in \eqref{S-matrix}, 
which serves as a clue to find the desired adjacency matrix.
(See Appendix for scattering theory required in this section.)
Using \eqref{S-matrix}, we find that
the following three graphs
\begin{equation}
\includegraphics[width=0.85\textwidth]{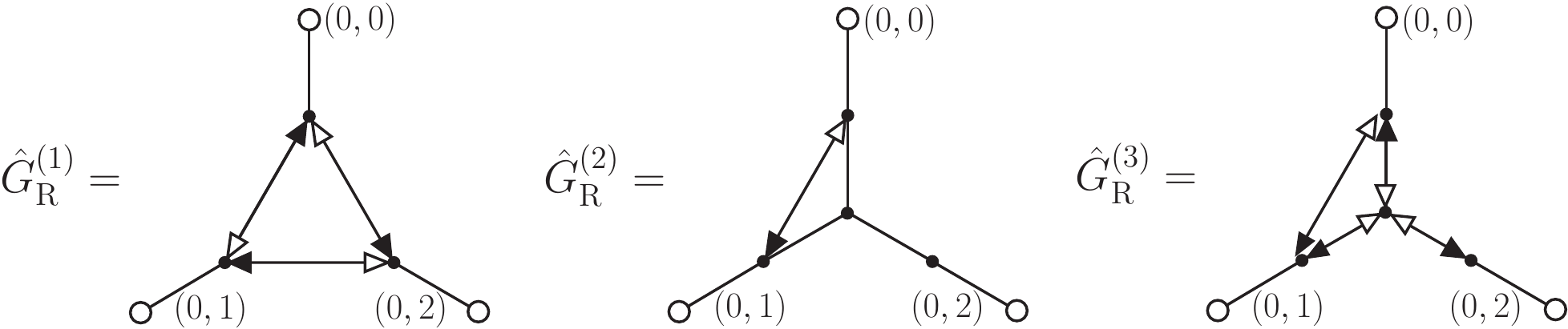}
\label{RA-graph}
\end{equation}
with the abbreviation of the directed weighted edge
\begin{equation}
\includegraphics[width=0.8\textwidth]{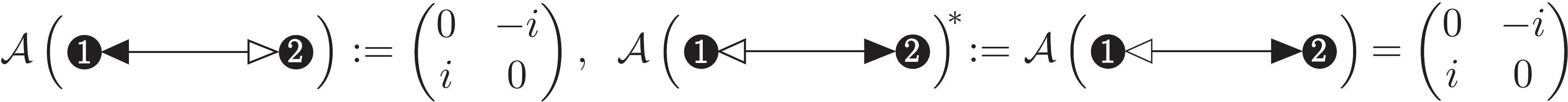}
\end{equation}
all implement the roundabout gate up to the sign (the minus sign 
is necessary for $\hat{G}_\mathrm{R}^{(1)})$,
when the momentum of the walker passing through them is exactly
$k=-\pi/2$:
$S(-\pi/2)=U_{\mathrm{R}}$ 
for the red walker and 
$\tilde{S}(-\pi/2)=S(-\pi/2)^{\mathrm{T}}=U^{\dagger}_{\mathrm{R}}$
(cf. \eqref{transpose}) for the blue walker.
More explicitly,  the matrix elements of $S(k)=\sum_{m,n}S_{mn}(k)
|0,m\ket\bra 0,n|$ 
are given by
\begin{alignat}{4}
&S_{00}(k)=\frac{-e^{4 i k}}{1+2i \tan k},\,\,\,&&
 S_{10}(k)=\frac{-2 e^{\frac{7 i k}{2}-
\frac{i \pi }{4}} \cos \left(\frac{k}{2}+\frac{\pi }{4}\right)}{2-i \cot k}, \,\,\,&&
S_{20}(k)=\frac{-2 i e^{\frac{7 i k}{2}-
\frac{i \pi }{4}} \sin \left(\frac{k}{2}+\frac{\pi }{4}\right)}{2-i \cot k}, \nn \\
&S_{11}(k)=S_{00}(k),\,\,\,&&
S_{21}(k)=S_{10}(k),\,\,\,&&
S_{22}(k)=S_{00}(k),
\label{S1}
\end{alignat}
for $\hat{G}^{(1)}_\mathrm{R}$ and
\begin{alignat}{4}
&S_{00}(k)=\frac{-e^{4 i k}}{1+2i \tan k},\,\,\,&&
 S_{10}(k)=\frac{2 e^{\frac{7 i k}{2}-
\frac{i \pi }{4}} \cos \left(\frac{k}{2}+\frac{\pi }{4}\right)}{2-i \cot k}, \,\,\,&&
S_{20}(k)=\frac{2 e^{\frac{9 i k}{2}-
\frac{i \pi }{4}} \sin \left(\frac{k}{2}+\frac{\pi }{4}\right)}{2-i \cot k}, \nn \\
&S_{11}(k)=S_{00}(k),\,\,\,&&
S_{21}(k)=\frac{2 e^{\frac{9 i k}{2}+
\frac{i \pi }{4}} \cos \left(\frac{k}{2}+\frac{\pi }{4}\right)}{2-i \cot (k)},\,\,\,&&
S_{22}(k)=e^{2ik}S_{00}(k),
\label{S2}
\end{alignat}
for $\hat{G}^{(2)}_\mathrm{R}$ and  $\hat{G}^{(3)}_\mathrm{R}$. The other elements are
determined by the relation 
$S(k)=S(-k)^{\dagger}$ which holds for $k\in\mathbb{R}$ (see \eqref{unitary}).

\begin{figure}[t]
\centering
\includegraphics[width=0.75\textwidth]{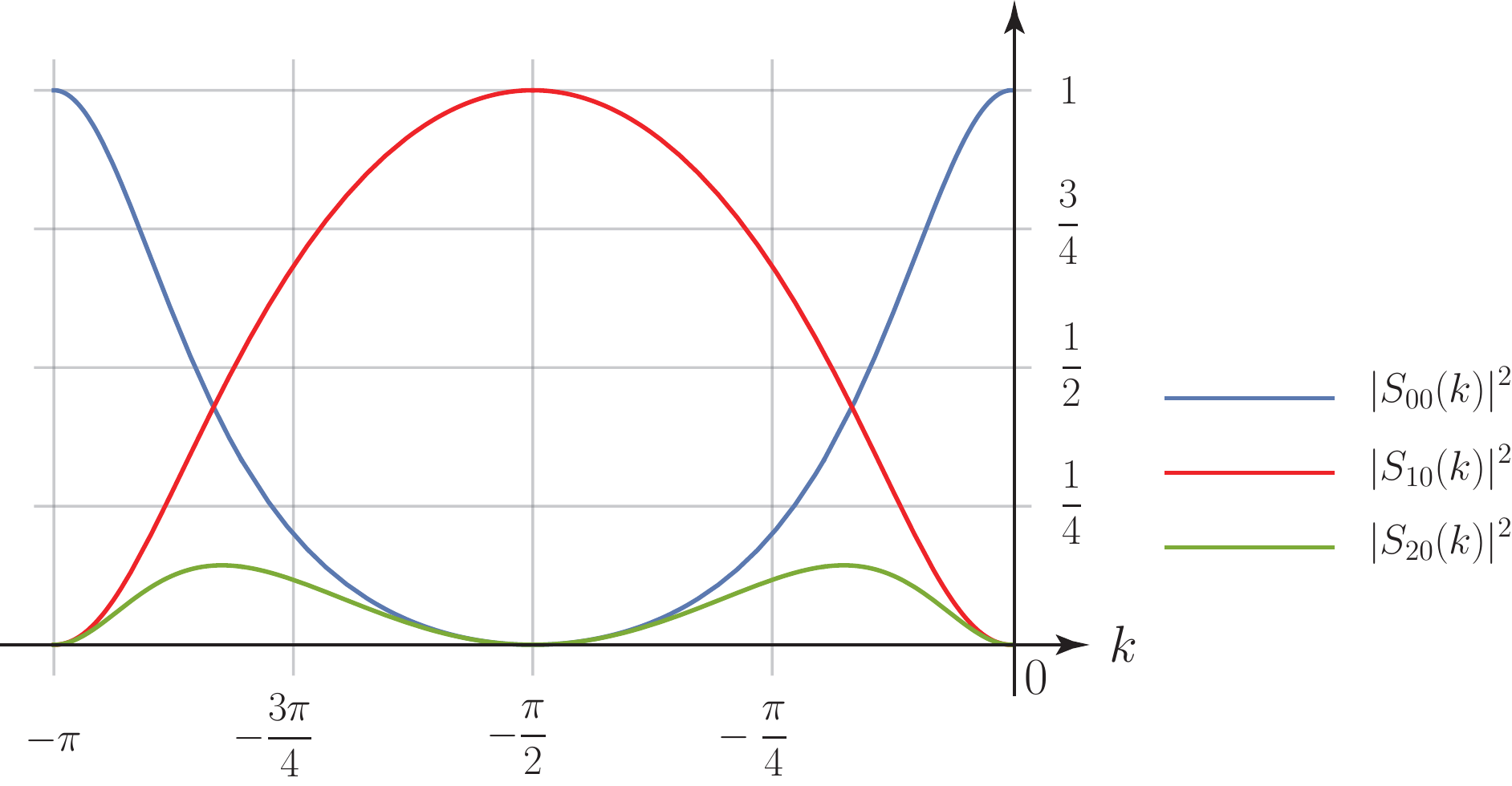}
\caption{The momentum dependence of the probability 
$|S_{m0}(k)|^2$ ($m=0,1,2$) that a walker entering $\hat{G}_\mathrm{R}^{(j)}$ ($j\in\{1,2,3\}$)
\eqref{RA-graph} from path $0$
will be found on path $m$ ($m=0,1,2$). The walker with momentum $k=-\pi/2$ completely
transmits from path $0$ to path $1$.}
\label{T-prob}
\end{figure}

The probability that a walker entering $\hat{G}_\mathrm{R}$ from path 
$m\in\mathbb{Z}/3\mathbb{Z}=\{0,1,2\}$
will be found on path $n\in\mathbb{Z}/3\mathbb{Z}=\{0,1,2\}$ is given by 
$|S_{mn}(k)|^2$, which is the same
for all three cases in \eqref{RA-graph}. We see that 
the walker with momentum $k=-\pi/2$ 
perfectly transmits from $n$ to $m=n+1$. In Fig.~\ref{T-prob}, 
the momentum dependence of 
the transmission probabilities are shown. In summary, 
the subgraphs $G^{(j)}_\mathrm{R}$ and  $G^{\ast (j)}_\mathrm{R}$ 
($j=1,2,3$) are, respectively, physical
implementations of the roundabout gates $U_\mathrm{R}^{(\mathrm{l})}$ 
and $U_\mathrm{R}^{(\mathrm{r})}$ 
for a quantum walker with momentum $k=-\pi/2$. Graphically,
\begin{equation}
 \includegraphics[width=0.4\textwidth]{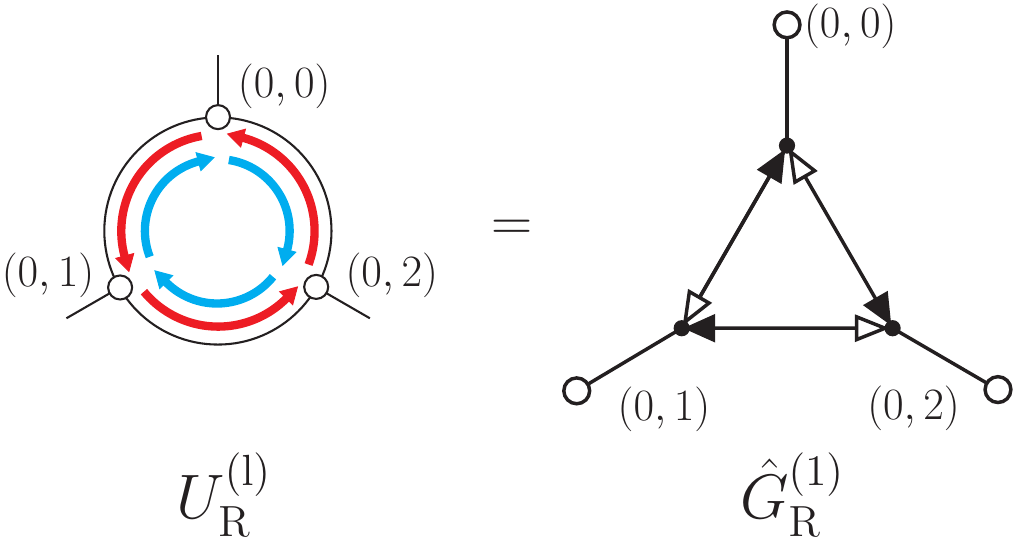},\quad
    \includegraphics[width=0.4\textwidth]{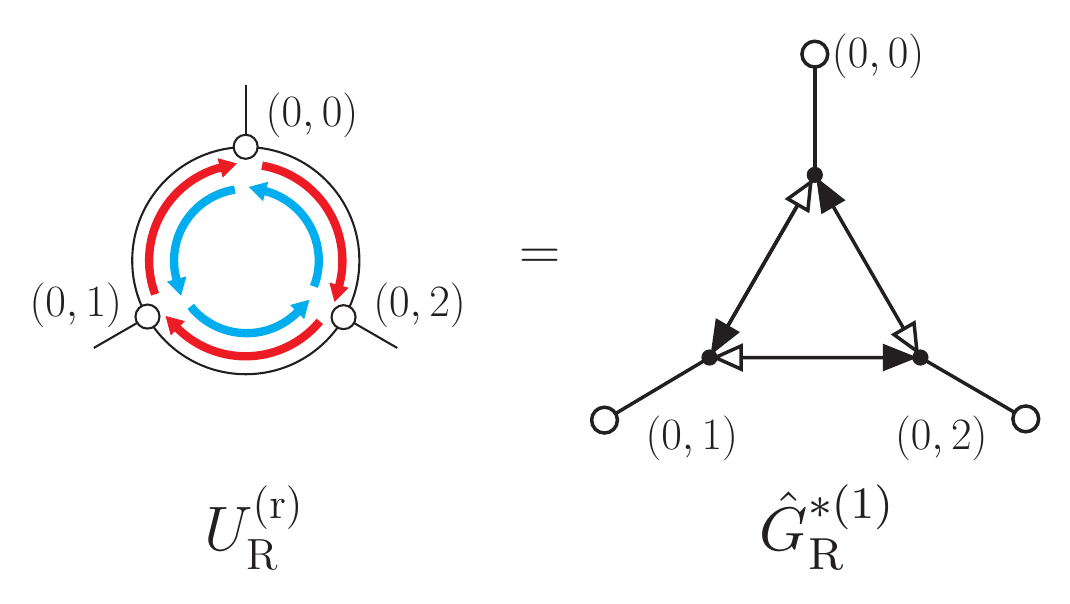}
\label{RA-GR}
\end{equation}
The same is true for $G^{(j)}_\mathrm{R}$ and 
$G^{\ast (j)}_\mathrm{R}$ ($j=2,3$) as above.
Note that $\hat{G}^\ast_\mathrm{R}$ is obtained by reversing all 
the arrows in $\hat{G}_\mathrm{R}$ (eq.~\eqref{RA-graph}),
because $\mathcal{A}(\hat{G}^{\ast}_\mathrm{R})
=\mathcal{A}(\hat{G}_\mathrm{R})^\ast$.

In the actual implementation,  
we use a wave packet \eqref{wave-packet} 
as a quantum walker and consider the scattering state 
on the subgraph $\hat{G}_{\mathrm{R}}$ with sufficiently long (but finite length) 
paths attached. 
The effective length of the subgraph $\hat{G}_{\mathrm{R}}$ 
can be evaluated by the non-trivial phase shift of the wave packet 
output from $\hat{G}_{\mathrm{R}}$  after scattering.
The wave packet $|\Psi_{\mathrm{out}},t\ket$ 
for the red walker exiting from $(0,l)$, which entered 
$\hat{G}_\mathrm{R}$ from $(0,j)$ (cf. \eqref{scattering-state} and 
\eqref{wave-packet}), is given by
\begin{align}
|\Psi_{\mathrm{out}},t\ket&=
\frac{1}{\sqrt{2\pi}}
\int_{k\simeq k_\mathrm{p}} dk S_{lj}(k)f(k) e^{i k x-i E(k) t}|0\ket_\mathrm{c}|x\ket\nn \\
&\simeq
\frac{S_{lj}(k_\mathrm{p})e^{ik_\mathrm{p} x-i E(k_\mathrm{p})t}}{\sqrt{2\pi}}\int_{k\simeq k_\mathrm{p}} dk f(k)
e^{i(k-k_\mathrm{p})(x-i(\log S_{lj})'(k_\mathrm{p})-E'(k_\mathrm{p})t)}|0\ket_\mathrm{c}|x\ket.
\end{align}
Thus, the effective length 
\begin{equation}
\ell_{lj}(k):=-i (\log S_{lj})'(k)
\end{equation}
 for the red walker is calculated 
by  \eqref{S1} or \eqref{S2}:
\begin{equation}
\ell_{lj}\left(-\frac{\pi}{2}\right)= 3\delta_{l,j+1} \text{ for $G_\mathrm{R}^{(1)}$},
\quad
\ell_{lj}\left(-\frac{\pi}{2}\right)=
\begin{cases}
3\delta_{l,j+1} &\text{for $j=0$} \\
4\delta_{l,j+1} &\text{for $j=1,2$}
\end{cases}
\text{ for $G_\mathrm{R}^{(2)}$ and  $G_\mathrm{R}^{(3)}$},
\label{eff-red}
\end{equation}
where $j,l\in\mathbb{Z}/3\mathbb{Z}=\{0,1,2\}$. On the other hand,
the effective length $\tilde{\ell}_{lj}(k)$ for the blue walker 
is given by the  relation
\begin{equation}
\tilde{\ell}_{lj}(k)=\ell_{jl}(k) \quad (k\in\mathbb{R}),
\label{eff-blue}
\end{equation}
which follows from \eqref{transpose}. 
These effective lengths are necessary for building actual 
finite-size quantum circuits, where a synchronization of the motion of 
quantum walkers becomes crucial. See Sec.~5 for details.

\section{Elementary quantum gates} \label{EG}

In this section, we describe how to implement a universal quantum 
gate set via two-level quantum walkers. As described in Sec.~\ref{qwalk}, 
information is spatially encoded by the positions of red quantum walkers.
Information processing is carried out by moving multiple quantum walkers to 
appropriate positions. In our architecture, the internal 
state of the quantum walker serves as a temporal storage medium
during this process. Any unitary transformation to a single-qubit 
state can be replaced by a transformation to the internal state of 
the quantum walker. There, the roundabout gate not only switches the 
position of the walkers but acts as an information encoder and decoder. 
In addition, combining two-particle scattering on an infinite path,
one can implement a controlled gate.

\subsection{Single-qubit gates}
First, let us implement a unitary operator
$U$ acting on a single-qubit state $|q_j\ket$ defined as \eqref{dual1}.
In our scheme, we express the state as the position of the red quantum walker:
\begin{equation}
|0\ket_\mathrm{c}|q_j\ket=|0\ket_\mathrm{c}
|2j+q_j\ket_\mathrm{p}\in \mathbb{C}^2\otimes \mathbb{C}^2.
\label{dual-internal}
\end{equation}
The information $|q_j\ket$ can be encoded into the internal state
of the walker passing through the following encoder $U_\mathrm{E}$
which consists of the Pauli X gate $X_\mathrm{c}$  
(eq. \eqref{fig-gate}) and the
roundabout gate  $U_\mathrm{R}^{(\mathrm{r})}$ (eq. \eqref{fig-RA}):
\begin{equation}
\begin{array}{c}
\includegraphics[width=0.7\textwidth]{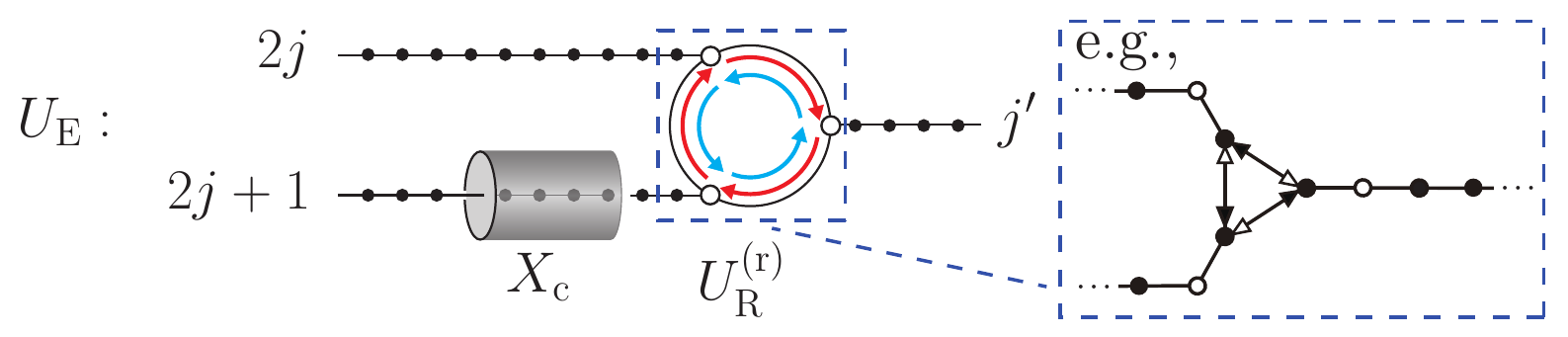}
\end{array}.
\label{encoder}
\end{equation}
Let us explain in detail. 
A red walker prepared on the $(2j)$th or $(2j+1)$th input path (possibly in 
superposition) moves toward the roundabout gate $U_\mathrm{R}^{(\mathrm{r})}$; the red walker on the $(2j+1)$th path changes color to blue 
at the $X_\mathrm{c}$ gate before entering $U_\mathrm{R}^{(\mathrm{r})}$. 
The roundabout gate $U_\mathrm{R}^{(\mathrm{r})}$ consists of the subgraph 
$\hat{G}_\mathrm{R}^{\ast}$ 
moves the red walker (resp. blue walker) clockwise 
(resp. counterclockwise) to output path $j'$, as shown in the 
previous section.
For example, the inset depicts $\hat{G}^{\ast (1)}_\mathrm{R}$ (cf. \eqref{RA-GR}).
The effective length of $U_\mathrm{R}^{(\mathrm{r})}$
can be exactly the same for the red and blue walkers,
if the terminal vertices $(0,j)$ $(j\in\{0,1,2\})$ 
of $\hat{G}_\mathrm{R}^{\ast}$ as in \eqref{RA-GR} and \eqref{RA-graph}
are connected to paths $\{2j, 2j+1,j'\}$ in \eqref{encoder} 
so that $(0,0)\to (0,2j)$, $(0,1)\to (0,2j+1)$ and $(0,2)\to (0,j')$.
More precisely, one finds
\begin{equation}
\ell_{j',2j}\left(-\frac{\pi}{2}\right)=\tilde{\ell}_{j',2j+1}
\left(-\frac{\pi}{2}\right)=
\begin{cases}
3 &\text{for $\hat{G}^{\ast (1)}_\mathrm{R}$}\\
4 &\text{for $\hat{G}^{\ast (2)}_\mathrm{R}$ and $G^{\ast (3)}_\mathrm{R}$ }\\
\end{cases},
\end{equation}
where $\ell_{j',2j}(-\pi/2)$ and $\tilde{\ell}_{j',2j+1}(-\pi/2)$ are,
respectively, the effective lengths of $U_\mathrm{R}^{(\mathrm{r})}$
for the red and blue walker. This is derived from \eqref{eff-red}, 
\eqref{eff-blue} and the fact that the $S$-matrix  for 
$\hat{G}^{\ast}_\mathrm{R}$ is given by the transpose of that for 
$\hat{G}_\mathrm{R}$, as shown in \eqref{transpose}.
Eq.~\eqref{encoder} equivalently reads
\begin{equation}
\eqref{dual-internal}  
\xmapsto{X_\mathrm{c}\otimes (|2j+1\ket\bra 2j+1|_\mathrm{p})} 
\delta_{q_j, 0}|0\ket_\mathrm{c}|2j\ket_\mathrm{p}+\delta_{q_j, 1}|1\ket_\mathrm{c}
|2j+1\ket_\mathrm{p}
\xmapsto{U_\mathrm{R}^{(\mathrm{r})}}
|q_j\ket_\mathrm{c} |j'\ket_\mathrm{p}.
\label{encoder2}
\end{equation}
Applying a unitary gate $U_\mathrm{c}$ to the internal state
of the walker moving in the right direction on path $j'$, 
and then applying the decoder $U_\mathrm{D}$ (eq.~\eqref{fig-encode/decode})
realized by 
the reverse operation of the encoder $U_\mathrm{E}$ 
(eq.~\eqref{encoder})
(i.e., $U_\mathrm{D}=U_\mathrm{E}^{\dagger}$),
we obtain the desired state $U|q_j\ket$:
\begin{equation}
\eqref{encoder2} \xmapsto{U_\mathrm{c}}
\left(U_\mathrm{c}|q_j\ket_\mathrm{c}\right) |j'\ket_\mathrm{p}
\xmapsto{U_\mathrm{D}=U_\mathrm{E}^{\dagger}}
|0\ket_\mathrm{c}(U|q_j\ket).
\end{equation}
Graphically, the single-qubit gate is depicted as
\begin{equation}
  \begin{array}{c}
  \includegraphics[width=0.78\textwidth]{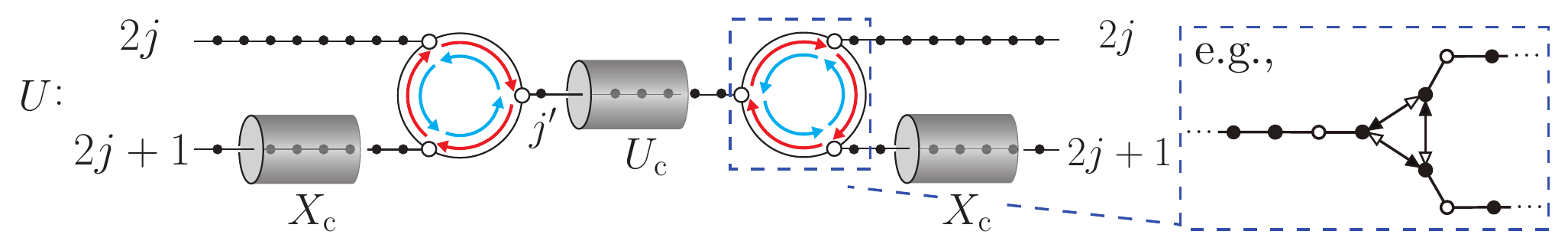}.
  \end{array}
\end{equation}

It is known \cite{nielsen2002quantum,williams2011quantum}
 that any single-qubit unitary gate $U$ 
is realized by 
\begin{equation}
U=e^{i\theta_0}R_z(\theta_1)R_y(\theta_2)R_z(\theta_3) \quad
(\theta_j\in\mathbb{R}),
\end{equation}
where 
\begin{equation}
R_y(\theta):=\exp\left(-i  \frac{\theta Y}{2}\right), 
\quad R_z(\theta):=\exp\left(-i \frac{\theta Z}{2}\right)
\end{equation}
are, respectively, the rotation operators 
about $y$ and $z$ axes of the Bloch sphere; $Y$ and $Z$ are
the Pauli matrices 
$\sigma_y$ and $\sigma_z$, respectively.
\begin{figure}[t]
\centering
\includegraphics[width=0.5\textwidth]{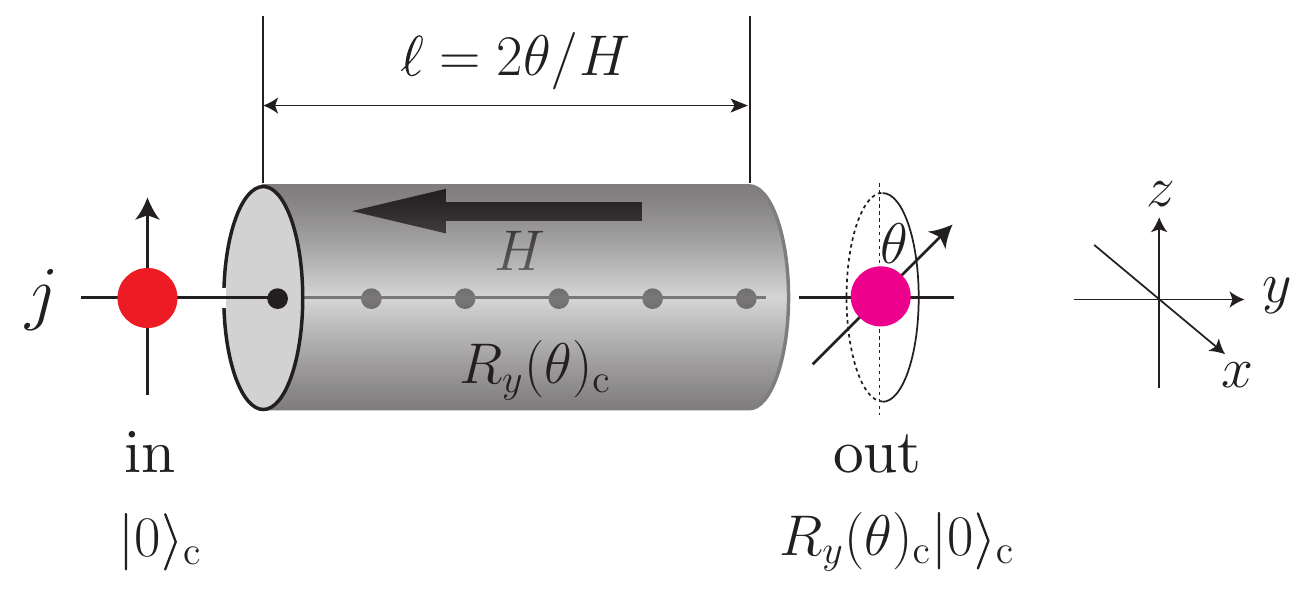}
\caption{A schematic description of the rotation gate
$R_y(\theta)_\mathrm{c}$ for the spin-1/2 fermionic system.  
A uniform magnetic field $H$ is applied in the negative direction
of the $y$-axis in the device set up like tunnels along the path.
Since the speed of the walker is $v_\mathrm{g}=2$ (eq.~\eqref{group})
and the Zeeman energy $\mathcal{H}_{\mathrm{ex}}=H\sigma_y/2=HY/2$,
$R_y(\theta)_\mathrm{c}=e^{-i\theta Y_\mathrm{c}/2}$ is 
implemented by just setting
the length of the device to $\ell=2\theta/H$.
}
\label{rotation}
\end{figure}
Actually, for instance, in the spin-1/2 fermionic system, the rotation 
gates acting on the spin state may be implemented by applying a uniform
magnetic field $H$ in a particular direction over a suitable interval of 
the path. 
See Fig.~\ref{rotation} for a schematic description of a 
rotation gate $R_y(\theta)_\mathrm{c}$ acting
on the internal state of the walker. 
Since the Zeeman energy $\mathcal{H}_\mathrm{ex}$
is given by $\mathcal{H}_\mathrm{ex}=H Y/2$ (resp.
$\mathcal{H}_\mathrm{ex}=H Z/2$) for the magnetic
field $H$ applied  in the negative direction of the 
$y$-axis (resp. $z$-axis), the spin state of the walker passing 
through the device over time $t$ is transformed by the
operator $R_y(H t)_\mathrm{c}=e^{-i t H Y_\mathrm{c}/2}$ (resp.  
$R_z(Ht)_\mathrm{c}=e^{-i tH  Z_\mathrm{c} /2}$). 
Because the (group) velocity $v_\mathrm{g}$
of the quantum walker with momentum $k=-\pi/2$ is
\begin{equation}
v_\mathrm{g}=E'\left(-\frac{\pi}{2}\right)=-2\sin\left(-\frac{\pi}{2}\right)=2
\label{group}
\end{equation} (see \eqref{energy}),
$R_{y,z}(\theta)_\mathrm{c}$ can be implemented by just setting the
device length $\ell$ so that 
$\theta=tH =\ell H/v_\mathrm{g}$, i.e., $\ell=2 \theta/H$.

\begin{figure}[t!]
  \centering
  \includegraphics[width=0.7\textwidth]{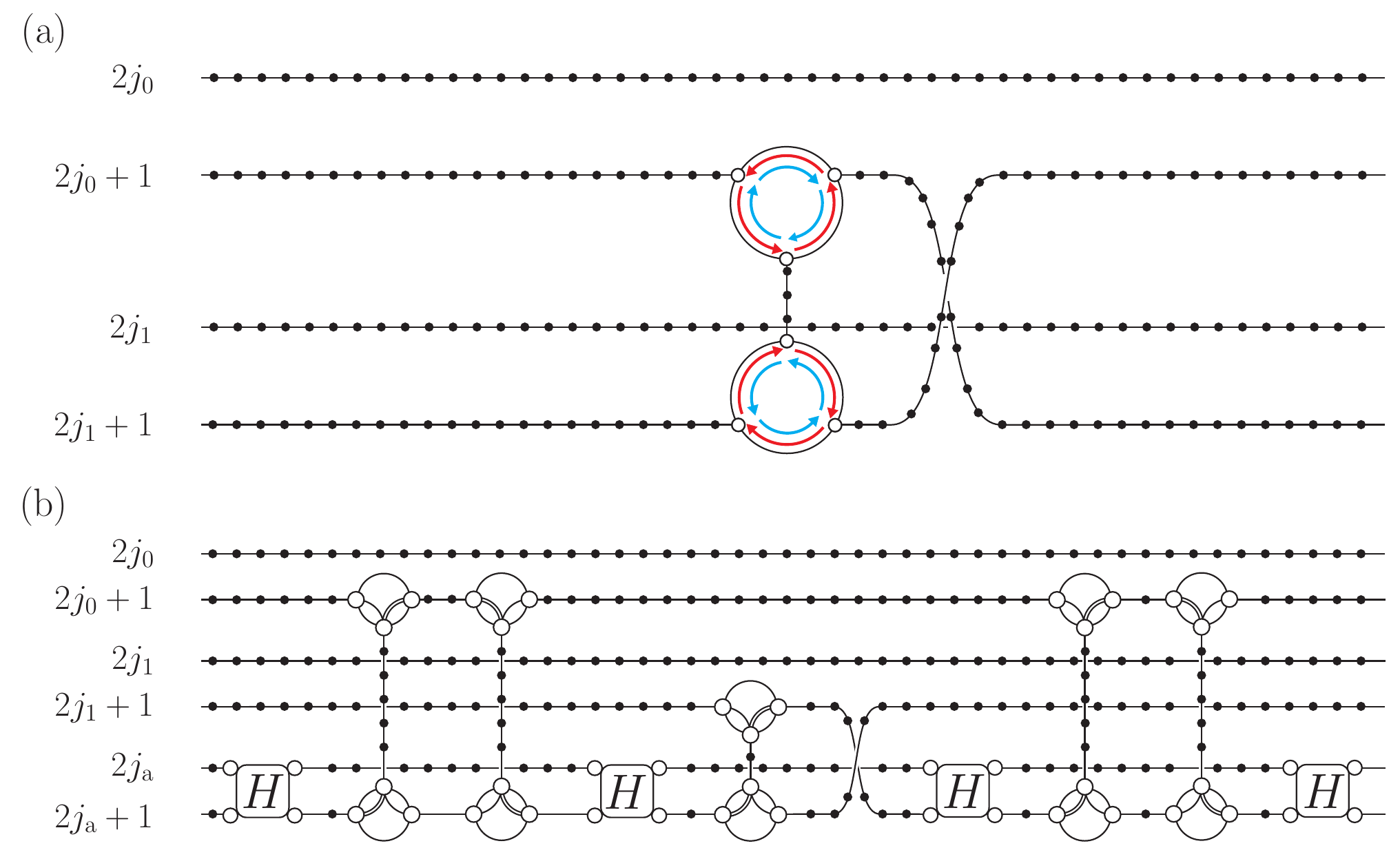}
  \caption{
  (a): A controlled-phase (CP) gate \eqref{CP-eq} implemented in  
  the present model. 
  (b): The CP gate implemented in the model 
  \cite{childs2013universal}. 
  In contrast to (b), single-particle scattering 
  is used  in (a) only for the implementation of the roundabout gate.
  Since (a) does not 
  require ancilla qubits (corresponding to 
  $|2j_\mathrm{a}\ket_\mathrm{p}$ and $|2j_\mathrm{a}+1\ket_\mathrm{p}$ in (b)),
  the architecture can be drastically simplified 
  compared to (b).
  }
  \label{CP}
  \end{figure}

\begin{figure}[t]
  \centering
  \includegraphics[width=0.82\textwidth]{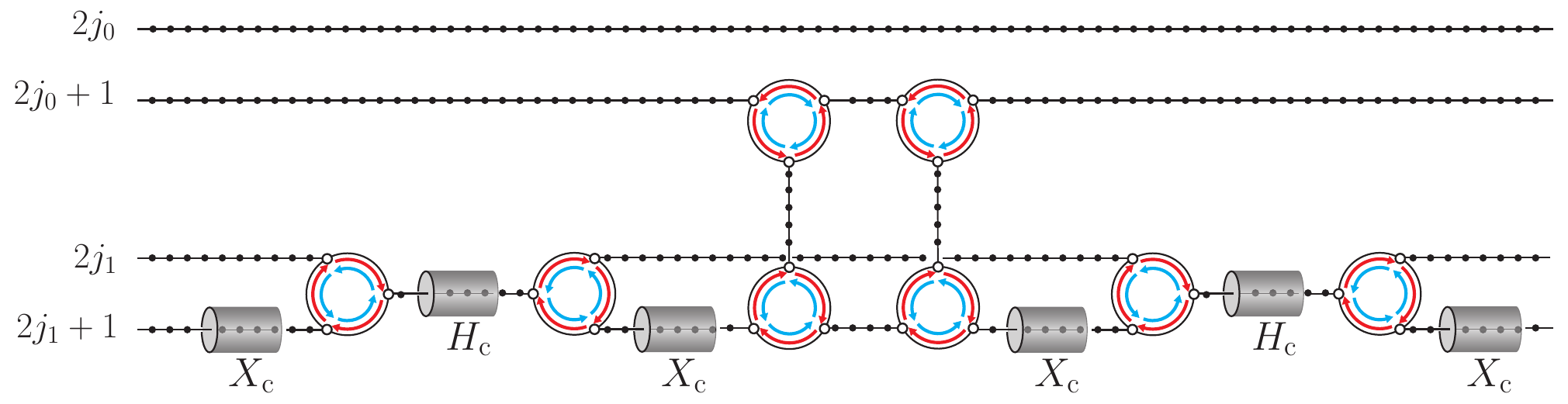}
  \caption{A graphical representation of the CNOT gate
  \eqref{CNOT}.
  }
  \label{CNOT-fig}
  \end{figure}

\subsection{Two-qubit gates}
Appropriately combining the scattering of two walkers with the 
same internal state and a single-qubit gate described above, 
we can implement a two-qubit gate. The aforementioned 
single-qubit rotation gates and a controlled-NOT (CNOT) gate 
are elements of a universal gate set \cite{nielsen2002quantum,
williams2011quantum}. In fact, the CNOT gate
$\mathrm{CNOT}_{j_0j_1}$ acting on a two-qubit state $|q_{j_1}\ket|q_{j_0}\ket$
is decomposed into
\begin{equation}
\mathrm{CNOT}_{j_0 j_1}=
H_{j_1}\mathrm{CP}^2_{j_0 j_1}H_{j_1},
\label{CNOT}
\end{equation}
where $H_{j_1}$ is the Hadamard gate on $|q_{j_1}\ket$ and
$\mathrm{CP}_{j_0 j_1}$ is a controlled phase (CP) gate
on $|q_{j_1}\ket |q_{j_0}\ket$:
\begin{equation}
\mathrm{CP}=
\begin{pmatrix}
1 & 0& 0& 0 \\
0 & 1& 0& 0 \\
0 & 0& 1& 0 \\
0 & 0& 0& \pm i 
\end{pmatrix}.
\label{CP-eq}
\end{equation}

As explained earlier, the action of  $H_{j_1}$  can be replaced by the action 
on the internal state of the quantum walker moving along path $j_1'$, 
which is connected to paths $2j_1$ and $2j_1+1$ by roundabout gates.
The CP gate, on the other hand,  is implemented by 
two-particle scattering on an infinite path: two particles, respectively 
moving along paths $2j_0+1$ and $2j_1+1$, are switched by the roundabout gate
to the same infinite path  and travel toward each other with momenta
$k_0=-\pi/2$ and $k_1=\pi/2$ to be scattered from each other.
Due to the conservation laws of energy and momentum, the individual momenta
are also conserved after scattering. As a result, only the global phase
of the two-particle wave function can be changed. In Appendix, we evaluate the phase
for both the bosonic and fermionic cases. For the bosonic system with
$u=\mp 4$ in the interaction term \eqref{interaction}, the wave function 
acquire a phase $S_{01}(k_0=-\pi/2,k_1=\pi/2)=\pm i$ after scattering (see
\eqref{boson-S}). 
For the fermionic case, a global phase  
$S_{01}(k_0=-\pi/2,k_1=\pi/2)=\pm i$ is acquired for $u=\mp 2$.
Consequently, the CP gate is implemented by two roundabout gates as depicted 
in Fig.~\ref{CP}. Characteristically, our architecture does not require any ancilla qubit;
therefore the structure of controlled gates can be significantly 
simplified, 
compared to those in \cite{childs2013universal}. Finally, we graphically 
represent the CNOT gate \eqref{CNOT} in Fig.~\ref{CNOT-fig}.

In the actual implementation of the two-qubit gates,  
we use  wave packets \eqref{wave-packet} 
as  quantum walkers and consider the scattering state 
on a sufficiently long (but finite length) paths.
The effective length $\ell(-\pi/2)$ for each walker in the 
two-particle scattering is obtained via \eqref{wave-function}, \eqref{boson-S} 
and \eqref{fermion-S}:
\begin{align}
\ell\left(-\frac{\pi}{2}\right)&=-i \del_{k_0}\log 
S\left(k_0=-\frac{\pi}{2},k_1=\frac{\pi}{2}\right)=
i \del_{k_1}\log S\left(k_0=-\frac{\pi}{2},k_1=\frac{\pi}{2}\right)\nn \\
&=
\begin{cases}
0 &\text{for boson}\\
-1/2&\text{for fermion}
\end{cases}.
\label{effective-two}
\end{align}

Finally, we would like to comment on the reversibility of our architecture.
Since the adjacency matrix of the graph used in the current architecture is 
a Hermitian matrix, the time evolution of the quantum walkers is of 
course described by a unitary operator. In this sense, quantum computation 
using the present model is as reversible as ordinary quantum computers.
However, simply reversing the motions of the walkers on the output paths (i.e., 
giving them the opposite momenta) is not enough to return them to their initial 
positions on the input paths. 
To return them to their original positions, 
one must not only reverse the motions of the walkers, but also replace the 
clockwise roundabout gates with counterclockwise roundabout gates, 
and vice versa. Alternatively,  we can also return them to the original 
positions by changing the colors of the walkers on the output paths from red (spin-up) 
to blue (spin-down), 
reversing the direction of all the magnetic fields applied in the single-qubit devices
(see Fig.~\ref{rotation}), 
and reversing the motions of the walkers (i.e., time reversal symmetry). 
\section{Building a circuit}
\label{building a circuit}
The quantum walker is effectively realized by a wave packet consisting 
of  plane waves with momentum close to a specific value of
$k\simeq k_\mathrm {p}$  (see \eqref{wave-packet}, for instance).
As explained earlier, in our architecture, $k_\mathrm{p}=-\pi/2$ is assigned to 
design  quantum gates. 
To build a quantum circuit for practical use, 
we must truncate the semi-infinite paths connected to each subgraph
and consider scatterings of wave packets on a finite-size graph (circuit).
There, the timing of the two-particle scatterings becomes crucial: 
the two-particle scatterings must occur at the proper vertical paths. 
Here, we explain how to arrange
the quantum gates to build a practical circuit,
according to the procedure developed in  \cite{childs2013universal}.
Also an error bound is estimated. 

\begin{figure}[t!]
  \centering
  \includegraphics[width=0.93\textwidth]{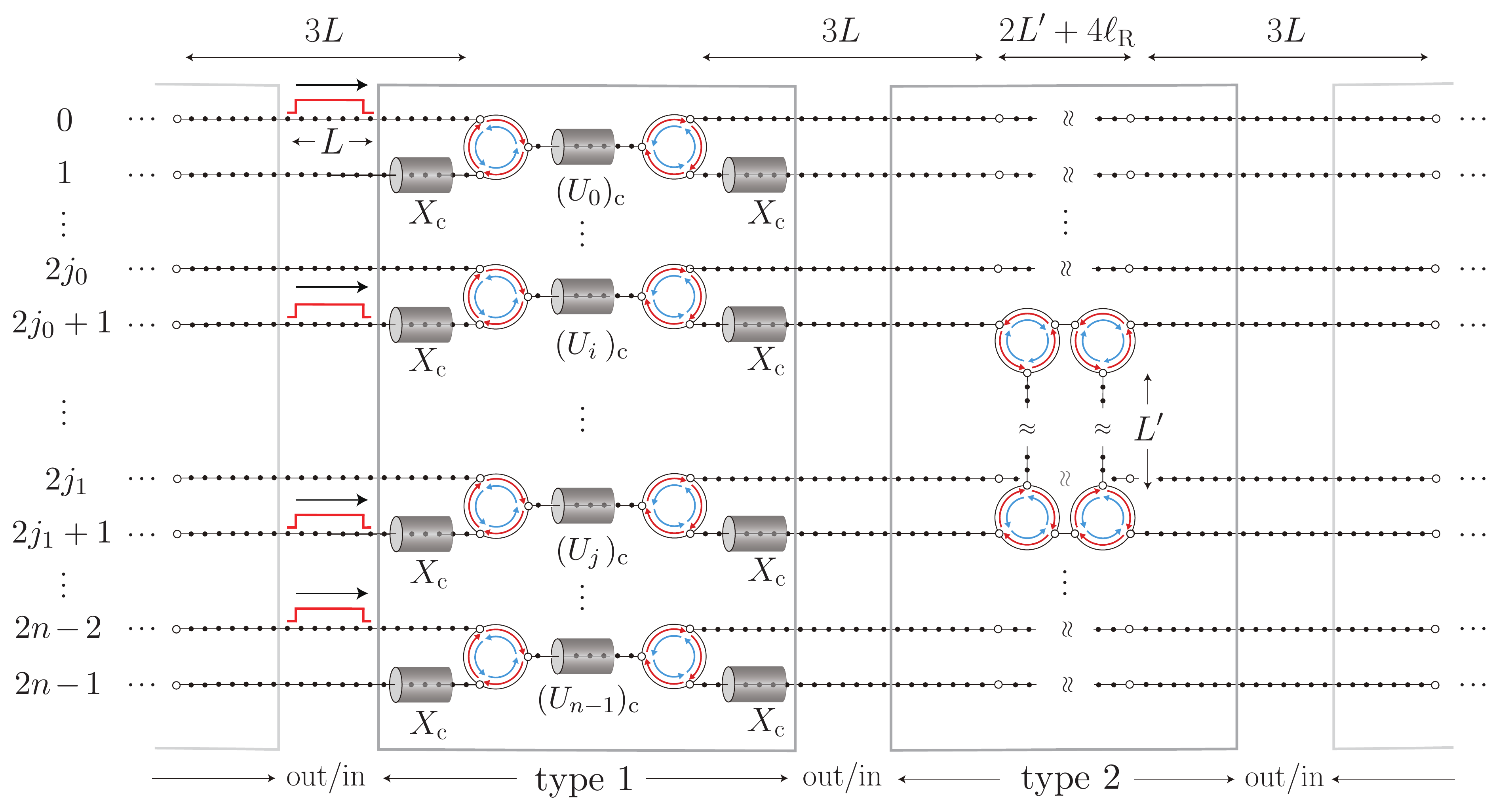}
  \caption{
In the actual $n$-qubit circuit, $n$ wave packets of 
length $L$ are employed  as quantum walkers to represent an $n$-qubit state. 
Any circuit can be built by 
a proper combination of type 1 and type 2 blocks. 
A block of type 1 consists of $n$ single qubit 
gates $U_0, \cdots, U_{n-1}$ (some of which may be identity gates).
A block of type 2 consists of a set of CP gates $\mathrm{CP}_{j_0 j_1}$ 
($j_0\ne j_1$, $j_0,j_1\in\{0,\cdots,n-1\}$) constructed by four roundabout gates. 
An identity gate in a block of type 2  is realized by simply connecting 
the input and output of the block with a straight path of length $2L'+4\ell_\mathrm{R}$,
where $L'\ge 4L$ is the length of each vertical path connecting the upper
and lower two roundabout gates within each CP gate, and $\ell_\mathrm{R}$
is the effective length of the roundabout gate (see \eqref{eff-red}).
Input/output paths 
of length $3L$ each are attached to each block, where
the output paths of one block are commonly used as 
the subsequent block. 
  }
  \label{blocks}
\end{figure}
Let us build an $n$-qubit circuit. Any quantum circuit can be designed 
by appropriately combining blocks consisting 
of single-qubit gates (``type 1 blocks") and blocks consisting of 
CP gates (``type 2 blocks"), as depicted in Fig.~\ref{blocks}.
The input/output 
$n$-qubit state
$|q_{n-1}\cdots q_0\ket$ ($q_j\in\{0,1\},\,\,j\in\{0,\cdots,n-1\}$)
of each block is dual-rail encoded  by $n$ red quantum walkers 
(cf. \eqref{dual-internal}),   given by 
the following wave packet of length $L$:
\begin{equation}
|q_j\ket=
\frac{1}{\sqrt{L}}
\sum_{x=L}^{2L-1} e^{\pm i\frac{\pi}{2} x}|0\ket_\mathrm{c}|x,2j+q_j\ket_\mathrm{in/out},
\label{input}
\end{equation}
where $|x,j\ket_\mathrm{in/out}$ denotes the position state of a particle on 
the $j$th input/output path
and the positive and negative signs in the exponential correspond to the input  and 
output wave packets, respectively.
Also, the length of each input/output path connecting to each block is set to $3L$,
and the output paths of one block are commonly used as the input 
paths for the subsequent block. The  state 
\eqref{input} is expressed as the wave packet
located in the center of each  path.
Incidentally, the input wave packet \eqref{input} itself is given by the
superposition of plane waves  \eqref{wave-packet} with the 
probability amplitude:
\begin{equation}
f(k)=\sqrt{\frac{2}{L\pi}}\frac{\sin\left(\frac{L-1}{2}(k+\frac{\pi}{2})\right)}
{k +\frac{\pi}{2}}
e^{\frac{3L-1}{2}\left(k+\frac{\pi}{2}\right)i}.
\label{weight}
\end{equation} 

Now let us consider how the quantum gates in blocks of type 1 and type 2 can 
be arranged so that the desired pair of red walkers can meet simultaneously 
on a particular vertical path.
As shown in Fig.~\ref{blocks}, a block of type 1 consists 
of $n$ single-qubit gates $U_0, \cdots, U_{n-1}$  (some of which may be identity gates), 
in which $(U_j)_\mathrm{c}$ acts on the internal state 
of the walker transferred from path $2j$ or $2j+1$ by the leftmost roundabout gate.
As explained in Sec.~4, it is possible to make the effective lengths of the roundabout 
gates for the red and blue walkers exactly the same by incorporating the 
graphs $\hat{G}_\mathrm{R}$ and $\hat{G}^{\ast}_\mathrm{R}$ in the proper 
orientation. 
If the quantum gates are assembled in such a manner, quantum walkers that 
simultaneously enter a block of type 1  will exit that block at the same time.
In other words, type 1 blocks do not affect the order of walkers.

A block of type 2 consists of a set of CP gates $\mathrm{CP}_{j_0 j_1}$ 
($j_0\ne j_1$, $j_0,j_1\in\{0,\cdots,n-1\}$) and identity gates.
In contrast to a block of type 1, an identity gate is realized by simply connecting 
the input and 
output with a straight path.
As a CP gate,  we employ
\begin{equation}
\mathrm{CP}_{j_0 j_1}=
\begin{pmatrix}
1 & 0& 0& 0 \\
0 & 1& 0& 0 \\
0 & 0& 1& 0 \\
0 & 0& 0& -1 
\end{pmatrix}_{j_0j_1}
\label{CP-eq2}
\end{equation}
instead of \eqref{CP-eq}, taking into account the effective length 
\eqref{effective-two} of the two-particle 
scattering of fermionic walkers .
This CP gate is  achieved by placing four roundabout gates as shown 
in Fig.~\ref{blocks}. The right terminal vertex on path $2j_0+1$ or $2j_1+1$ of the 
left roundabout gate can be directly connected to the right terminal vertex of the 
right roundabout gate without compromising the function of the gate.
Let us set the length of each vertical path to $L'\ge 4L$ and denote the effective length
of each roundabout gate by $\ell_\mathrm{R}$ (i.e., $\ell_\mathrm{R}=3$ or 
$\ell_\mathrm{R}=4$ from \eqref{eff-red}). 
Then, if we set the length of the straight path of the identity gate to 
$2L'+4\ell_\mathrm{R}$, a particular pair of red walkers simultaneously 
entering the block can be scattered within a specific vertical path.
In addition,
for the fermionic walks, we must take into account the effective length 
\eqref{effective-two}
 of the two-particle scattering: the scattered
fermionic walkers in the CP gate move ahead of other walkers by one 
vertex. In this case,  by increasing the number of vertices in each horizontal 
and vertical path by one in the next type 2 block, the desired two-particle 
scattering can occur within a specific vertical path in the next block.
Namely, in the $m$th type 2 block, the number of vertices in each horizontal 
and vertical path should be increased by $m$ more than in the first.

As explained in Sec.~3 and 4, the roundabout gate and the CP gate are 
designed based on  single- and two-particle scattering states of
plane waves with a definite momentum $k=k_\mathrm{p}$, which are
defined on the subgraph with semi-infinite path and on the 
straight vertical path of infinite length, respectively.
Therefore, to ensure the reliability of the architecture, it is also 
essential to estimate errors that
occur in an actual quantum circuit consisting of a 
finite-size graph on which the wave packets move as quantum
walkers.  Let 
$|\Psi_\mathrm{in}\ket$ be an input state in superposition (cf. \eqref{input}):
\begin{equation}
|\Psi_\mathrm{in}\ket = 
\frac{1}{\sqrt{L}}\sum_{\{q_j\}} \bigotimes^{n-1}_{j=0} \left(\sum_{x=L}^{2L-1} 
e^{i\frac{\pi}{2}x}|0\ket_\mathrm{c}|x, 2j+q_j\ket_\mathrm{in}  \right),
\end{equation}
and $e^{-i\mathcal{H}_\mathrm{tot}\tau}|\Psi_\mathrm{in}\ket$ be the 
output state after processing over time $\tau$ in a circuit
intended to implement a unitary operator $U$, 
where  $\mathcal{H}_\mathrm{tot}$ is the total Hamiltonian for
the circuit, consisting of the kinetic term \eqref{kinetic term}, 
interaction term \eqref{interaction}, and
Zeeman energies used in the single-qubit gates (see Sec.~4).
An error bound between the desired state $|\Psi_\mathrm{out}\ket$:
\begin{equation}
|\Psi_\mathrm{out}\ket = \frac{1}{\sqrt{L}}\sum_{\{q_j\},\{q_j'\}} 
\bra q'_{n-1}\cdots q_0'| U | q_{n-1}\cdots q_0\ket  \bigotimes^{n-1}_{j=0}
\left(\sum_{x=L}^{2L-1} e^{-i\frac{\pi}{2}x}|0\ket_\mathrm{c} |x, 2j+q^\prime_j\ket_\mathrm{out}\right),
\end{equation}
and the actual output state $e^{-i\mathcal{H}_\mathrm{tot}\tau}|\Psi_\mathrm{in}\ket$
is estimated by directly applying the method in \cite{childs2013universal}
(see also \cite{farhi2007quantum}). It yields
\begin{equation}
    \left\|\left|\Psi_\mathrm{out}\right\ket - 
e^{-i\mathcal{H}_\mathrm{tot}\tau} \left|\Psi_\mathrm{in}\right\ket
\right\| = O \left(gn^3 L^{-\frac14} \right),
\label{error}
\end{equation}
where $g$ is the total number of type 1 and type 2 blocks. 
This error bound is obtained by multiplying an error bound $O(g n L^{-1/4})$ 
due to the use of wave packets by an error bound $O(n^2)$ 
due to truncation of the semi-infinite paths in each block 
\cite{childs2013universal}. 
Note that errors in the gates acting on the internal states $|c\ket_\mathrm{c}$ 
can be neglected for the following reasons.
These gates, for instance,  $X_\mathrm{c}$ and $H_\mathrm{c}$ 
as in \eqref{Pauli} or \eqref{fig-gate} are designed for a wave packet
with group velocity $v_\mathrm{g}=2$ (see Fig.~\ref{rotation}). Errors in these gates 
mainly come from a distortion of the wave packet by higher order dispersion effect:
the velocity of the wave packet is corrected as
\begin{align}
v\simeq E'(k_\mathrm{p})+\frac{1}{2}(k-k_\mathrm{p})E''(k_\mathrm{p})+\frac{1}{6}(k-k_\mathrm{p})^2
E'''(k_\mathrm{p})
=v_\textrm{g}+O\left(\frac{1}{L^2}\right),
\label{correction}
\end{align}
where we have used $k_\mathrm{p}=-\pi/2$, $E(k)=2\cos k$ and also 
used $|k-k_\mathrm{p}|=O(1/L)$ which follows from \eqref{weight}.
Thus, the error caused by the correction  \eqref{correction} 
is estimated to be $O(n L^{-2})$  
at each block, which is negligible compared to the above 
error bound $O(n L^{-1/4})$ due to scatterings of wave packets. 

As pointed out in 
\cite{childs2013universal}, the error bound estimated in \eqref{error} is 
almost surely not optimal: a significant improvement can be expected.
Nevertheless, by setting $L=O(g^4n^{12})$, one finds that
universal quantum computation of arbitrary 
precision can be achieved with polynomial overhead. Namely, the total
number of vertices in the architecture is $O(g^5 n^{13})$
and the total computational time is $O(g^5 n^{12})$, which are formally the
same as the architecture in \cite{childs2013universal}. However, a feature of 
our architecture is that the number of blocks $g$ can be significantly 
reduced by using $O(n g)$ gates acting on the internal states, as shown 
in Fig.~\ref{CP}, which is expected to enable more efficient implementation 
for universal computation than in \cite{childs2013universal}.

\section{Summary and Discussion}
We have proposed a model of universal quantum computing using a 
multi-particle continuous-time quantum walk. Quantum information is 
dual-rail encoded by the quantum walkers with two internal states
(colors)
$|0\ket_\mathrm{c}$ (red) and $|1\ket_\mathrm{c}$ (blue).  To process the 
information spatially,
we have newly developed the roundabout gate that moves the quantum
walker from one path to the next, clockwise or counterclockwise, 
depending on the color of the quantum walker.
Any single-qubit unitary transformation is converted 
to a transformation for the color of the quantum 
walker. In this transformation, the roundabout gate acts as an
information encoder and decoder. Two quantum walkers can be scattered 
from each other on an infinite path to change a global phase of the 
two-particle wave function. An appropriate combination of a single-qubit 
gate and two-particle scattering yields a two-qubit controlled gate.

Our approach has several advantages. (i) Two-level quantum walkers can be realized 
with very mundane particles such as electrons and spin-$1$ bosons.
The Bose-Hubbard model (bosonic walks) and the extended Hubbard model (fermionic walks) 
employed in the architecture are models commonly used in many-particle physics.
(ii) Simplification of design is possible. The single-particle scattering is 
only applied to implement the roundabout gate.
Instead, a single-qubit gate is realized by a quantum device that acts on 
the color of the quantum walker.
The subgraph $\hat{G}_\mathrm{R}$ for the roundabout gate can be 
designed to be as simple as possible: the total number of vertices is at most 7, the 
maximum degree of $\hat{G}$ is three, and 
the edge weights of the graph take only $1$ and $\pm i$
(see \eqref{RA-graph} and \eqref{RA-GR}). 
The colors of the walker coexist temporarily  within 
the individual single-qubit gates.  In other words, maintaining the 
coherence of the colors is only required within individual single-qubit gates.
The two-particle scattering necessary to realize a two-qubit gate 
occurs only between a pair of red walkers and only in a straight path.
For the implementation, any ancilla qubit is unnecessary.
(iii) An automatic quantum computation is possible: the computation is 
done by just passing quantum walkers through appropriately designed paths.
(iv) A unified design of quantum computing compatible with an automatic 
qRAM is possible.

\begin{figure}[t]
\centering
\includegraphics[width=0.6\textwidth]{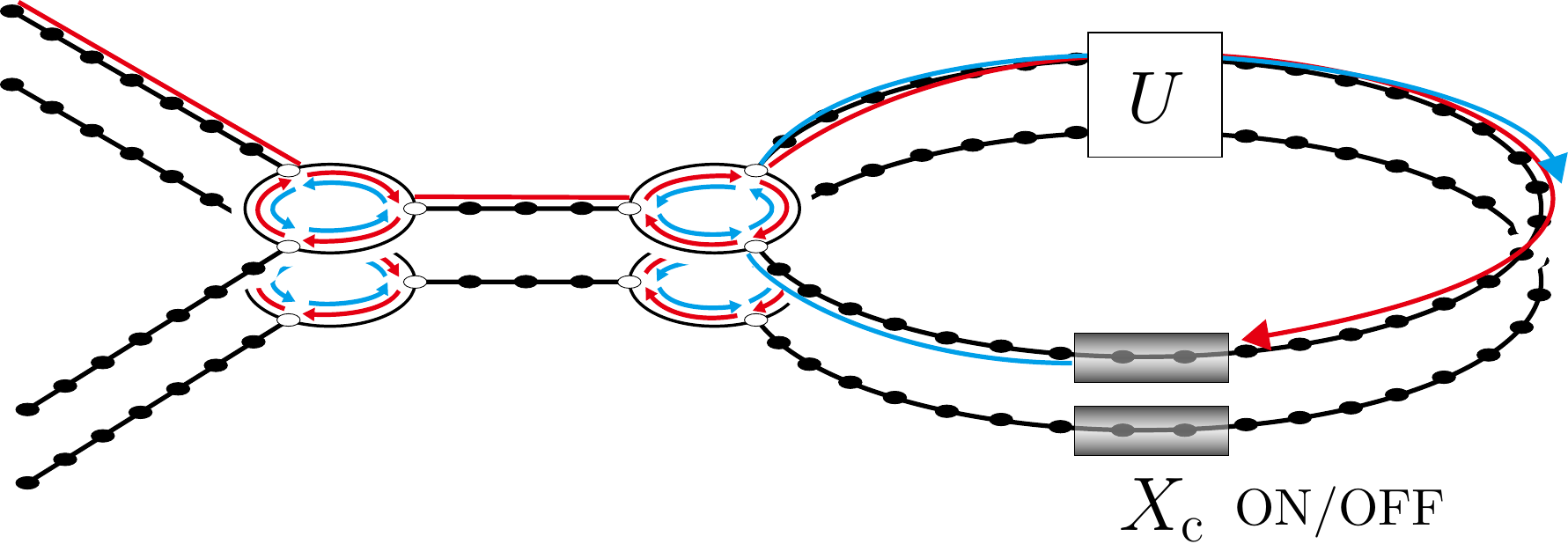}
\caption{An example of a circuit that serves as a quantum memory.
Information can be stored and retrieved at will by switching on 
the Pauli X gates installed on the loop.
Iterative calculations can also be easily carried out by simply 
placing unitary gates on the loop, which may be helpful to perform, 
for instance, the Grover algorithm.}
\label{qM}
\end{figure}

Here, we would like to discuss some possible applications using our 
architecture.  A crucial and non-trivial application is a physical 
realization of a qRAM, as briefly introduced in the introduction
and above. 
The details will be proposed in the next paper \cite{asaka2021two}.
Some simple quantum memory can also be constructed by a
suitable combination of the roundabout gate, as in 
Fig.~\ref{qM}. The quantum walkers, which represent the 
result of the calculation, can be led into the loops by the roundabout gates 
and stored as information. The information can be freely retrieved by 
switching on the Pauli X gates installed on the loops. Iterative calculations can also be easily carried out by simply 
placing unitary gates on the loop. It might be helpful to efficiently perform
information processing such as the Grover search \cite{grover1996fast}, 
the quantum phase estimation \cite{shor1994algorithms} and the quantum 
version of fast Fourier transform \cite{asaka2020quantum}.

The roundabout gate is essential in the present architecture and  automatic 
qRAM. Theoretically this can be realized by imposing an internal-state-dependent
gauge factor as shown in \eqref{kinetic term}.
Experimentally, it may be possible to achieve this, e.g., by applying
the Aharonov-Kasher effect \cite{aharonov1984topological,zvyagin1992aharonov}, 
but this still remains open.

\section*{Acknowledgment}
%
The present work was partially supported by Grant-in-Aid for Scientific
Research (C) No. 20K03793 from the Japan Society for the 
Promotion of Science.

\begin{appendix}
\section{Single- and two-particle scattering}
\label{scattering}
In this appendix, to make our paper self-contained, 
we summarize single- and two-particle scattering states 
of indistinguishable quantum walkers. 
\subsection{Single-particle scattering on a graph}
First, we consider the single-particle scattering on the graph 
$G$ defined in Fig.~\ref{graph-G}, using the procedure introduced in 
\cite{childs2013universal,childs2012levinson}.
Since the color of a single walker does not change
(see \eqref{kinetic term}), the single-particle scattering can be 
considered independently in the case of a red walker and 
a blue walker. Therefore, first, we fix the color of the walker to red 
($c=0$) (correspondingly, we only consider the
part of $c=0$ in the kinetic term \eqref{kinetic term}),
and omit the internal state $|0\ket_\mathrm{c}$ to simplicity
the notation. Later we will show that the single-particle $S$-matrix 
of the blue walker is nothing but the transpose
of that for the red walker.

Consider the process in which an incident walker (wave packet) 
with a momentum near a specific value of $k \in (-\pi,0)$
passes through the $j$th semi-infinite path toward the graph $\hat{G}$
and exits to some semi-infinite paths $l$ (in superposition) after 
scattering from $\hat{G}$. 
Since an arbitrary wave packet is given by the superposition of 
plane waves of momentum close to $k$ as in  \eqref{wave-packet}, the scattering 
state $|\varphi_j(k)\ket$ with a definite $k$ gives us information 
about how the wave packet scatters from $\hat{G}$.
It is generally written as
\begin{equation}
|\varphi_j(k)\ket=\sum_{x=0}^{\infty}e^{-ik x}|x,j\ket
+\sum_{l=0}^{N-1} \sum_{x=0}^\infty S_{lj}(k)e^{ik x}|x,l\ket+
\sum_{x\in V(\hat{G})\setminus T} \psi_j(x;k)|x\ket,
\label{scattering-state}
\end{equation}
where $|x,j\ket$ denotes a walker at $x$ on the $j$th semi-infinite
path (i.e. the walker at $(x,j)$),
$V(\hat{G}) \setminus T$ is the set of the $M$ internal vertices
of $\hat{G}$,
$S_{lj}(k)\in \mathbb{C}$ is the element of the $S$-matrix,
and $\psi_j(x;k)\in \mathbb{C}$ is the wave-function 
on $\hat{G}$, which can be determined by the Schr\"odinger
equation
\begin{equation}
\bra x|\mathcal{K}_G|\varphi_j(k)\ket=E(k) \bra x|\varphi_j(k)\ket \quad (x\in V(G)).
\label{SE}
\end{equation}
Here, $\mathcal{K}_G$ is the kinetic term defined as 
\eqref{kinetic term}.
Note that the interaction term \eqref{interaction} does not contribute to the 
single-particle scattering, and the  color of the  walker is implicitly assumed 
to be red ($|0\ket_\mathrm{c}$) as explained above.
In particular, for $(x,l)$ ($x\ge 1$), we have
\begin{equation}
\bra x,l|\mathcal{K}_G| \varphi_j(k)\ket=
2\cos k \bra x,l|\varphi_j(k)\ket,\quad
\bra x,l|\varphi_j(k) \ket= e^{-ik x}\delta_{l,j}+
e^{ik x}S_{lj}(k),
\end{equation}
which determines the energy $E(k)$:
\begin{equation}
E(k)=2\cos k.
\label{energy}
\end{equation} 
On the other hand,
the
Schr\"odinger equation \eqref{SE} for $x\in \hat{G}$ is 
written as
\begin{equation}
\begin{pmatrix}
A & B^{\dagger} \\
B& D
\end{pmatrix}
\binom{I_N+S(k)}{\psi(k)}+
\binom{e^{-i k }I_N+e^{ik } S(k)}{0}=E(k)
\binom{I_N+ S(k)}{\psi(k)},
\label{schroedinger}
\end{equation}
where 
\begin{align}
&S(k):=\sum_{l=0}^{N-1}\sum_{j=0}^{N-1}S_{lj}(k)
|0,l\ket\bra 0,j|\in 
\End(\mathbb{C}^N), \nn \\
&\psi(k):=\sum_{x\in V(\hat{G})\setminus T}\sum_{j=0}^{N-1}
\psi_j(x;k)|x\ket\bra 0,j|
\in \Hom(\mathbb{C}^N,\mathbb{C}^M),
\end{align}
and the matrix consisting of 
$A\in \End(\mathbb{C}^N)$, $D\in\End(\mathbb{C}^M)$, and 
$B\in\Hom(\mathbb{C}^N,\mathbb{C}^M)$ is the
adjacency matrix $\mathcal{A}(\hat{G})$:
$A=A^{\dagger}$ and $D=D^{\dagger}$ are the adjacency matrices of
the $N$ terminal vertices and the $M$ internal vertices, respectively, and
$B$ is the adjacency matrix between the terminal  and the 
internal vertices.
Solving \eqref{schroedinger},  one straightforwardly finds
\begin{equation}
S(k)=-e^{2ik}Q^{-1}(k)Q(-k),\quad
Q(k):=\left\{1-e^{ik}\left(A+B^{\dagger}\frac{1}{2\cos k-D}B\right)\right\}.
\label{S-matrix}
\end{equation}
For  $k\in\mathbb{R}$, the $S$-matrix satisfies
\begin{equation}
S(k)^{\dagger}S(k)=S(-k)S(k)=1 \quad (k\in\mathbb{R}),
\label{unitary}
\end{equation}
which can be easily seen by noting that  $A$ and $D$ are Hermitian,
$Q(k)^{\dagger}=Q(-k)$, and $[Q(k),Q(-k)]=0$.

Finally, let us show that the single-particle $S$-matrix  
of the blue walker ($c=1$) (denote it by $\tilde{S}(k)$) is given by 
$\tilde{S}(k)=S(k)^{\mathrm{T}}$ for $k\in \mathbb{R}$, which is the 
transpose of the $S$-matrix of the red walker. By property \eqref{kinetic-adjacency}, 
$\tilde{S}(k)$ can be obtained by simply replacing 
$\mathcal{A}(\hat{G})$ with $\mathcal{A}(\hat{G})^{\ast}$.
More explicitly, it is  given  by replacing 
$\{A,B,D\}$ with  $\{A^{\ast}, B^{\ast}, D^{\ast}\}$ in \eqref{S-matrix}.
This replacement changes $S(k)$  to $S(-k)^{\ast}$ ($k\in\mathbb{R}$). 
Using $S(k)^{\dagger}=S(-k)$ ($k\in\mathbb{R}$) derived from \eqref{unitary}, we arrive at
\begin{equation}
\tilde{S}(k)=S(-k)^{\ast}=(S(k)^{\dagger})^\ast=S(k)^{\mathrm{T}}.
\label{transpose}
\end{equation}

\subsection{Two-particle scattering on an infinite path}

Next, we consider the two-particle scattering on an infinite 
path: the walker (wave packet) with a momentum close to a 
specific value of $k_0\in(-\pi,0)$ moves right (down) toward the
walker which moves left (up) with a momentum close to $k_1\in(0,\pi)$.
 Since, in our architecture, we are concerned with 
the scattering of red walkers, here, we only consider the spatial part of the states
as in the single-particle scattering.
The two-particle scattering process can be characterized by the 
scattering state  $|\psi(k_0,k_1)\ket$ with
definite values of $(k_0,k_1)$. More explicitly, 
\begin{equation}
|\psi(k_0,k_1)\ket=\sum_{x_0,x_1\in\mathbb{Z}}
\psi(x_0,x_1;k_0,k_1)|x_0,x_1\ket,
\end{equation}
where $x_0, x_1\in\mathbb{Z}$ are vertices on
an infinite path and
the wave function $\psi(x_0,x_1;k_0,k_1)$ is written as
\begin{equation}
\psi(x_0,x_1;k_0,k_1)=\begin{dcases}
e^{i k_0 x_0+i k_1 x_1}\pm S_{01}(k_0,k_1)
 e^{ik_1 x_0+i k_0 x_1} \,\, &(x_0<x_1) \\
S_{01}(k_0,k_1) e^{ik_0 x_0+i k_1 x_1}
\pm e^{i k_1 x_0+i k_0 x_1} \,\, &
(x_0>x_1)
\end{dcases}.
\label{wave-function}
\end{equation}
$S_{01}\in\mathbb{C}$ is the scattering amplitude and the
$+/-$ sign corresponds to bosons/fermions. For the bosonic
case, $\psi(x,x)$ is given by putting formally $x_0=x_1=x$
in the above equation.

The form of the wave function 
\eqref{wave-function} is followed by the following facts:
\begin{enumerate}
 \item The
individual momenta $k_0$ and $k_1$ are conserved
after the scattering, due to the conservation law of the total 
energy $2(\cos k_0+\cos k_1)$ and momentum $k_0+k_1$. 
The two walkers only acquire a phase factor $S_{01}(k_0,k_1)$
after the scattering. 
\item The wave function is symmetric/antisymmetric 
under particle exchange, reflecting the Bose/Fermi statistics.  
\item Interactions are limited to on-site or nearest-neighbor pairs. 
\end{enumerate}
Indeed, for $x_0\ne x_1$ (resp. $x_0\ne x_1+1$) in 
the bosonic (resp. fermionic) case,
one can easily check that 
the wave function \eqref{wave-function} satisfies the Schr\"odinger equation 
\begin{equation}
\bra x_0, x_1| \mathcal{H}_G|\psi(k_0,k_1) \ket=2(\cos k_0+\cos k_1) 
\psi(x_0,x_1;k_0,k_1),
\label{schroedinger2}
\end{equation}
where $\mathcal{H}_G$ (eq.~\eqref{Hamiltonian}) with the interaction term \eqref{interaction} is
defined on an infinite path, i.e., $\theta_{jk}=0$.
Below, solving \eqref{schroedinger2} at $x_0=x_1$ (resp. $x_0=x_1-1$) for 
the bosonic (resp. fermionic) case, we determine the scattering amplitude
$S_{01}(k_0,k_1)$.
\subsubsection{Bosonic walkers}
We adopt the on-site  (Bose-Hubbard) interaction \eqref{interaction} 
for the bosonic quantum walkers. The LHS of the 
Schr\"odinger equation \eqref{schroedinger2} at $x_0=x_1=:x$
gives
\begin{align}
\text{LHS}&=
\psi(x,x+1;k_0,k_1)+\psi(x-1,x;k_0,k_1) \nn \\
&\quad +\psi(x,x-1;k_0,k_1)+\psi(x+1,x;k_0,k_1)+
u \psi(x,x;k_0,k_1) \nn \\
&=\left[2(e^{ik_1}+e^{-ik_0})+u+S_{01}(k_0,k_1)
\left\{2(e^{ik_0}+e^{-ik_1})+u\right\}\right]e^{i(k_0+k_1)x}.
\end{align}
Because the RHS of \eqref{schroedinger2} is given by 
\begin{equation}
\text{RHS}=2(\cos k_0+\cos k_1)
(1+S_{01}(k_0,k_1))e^{i(k_0+k_1)x},
\end{equation}
 we arrive at
\begin{equation}
S_{01}(k_0,k_1)=\frac{2(\sin k_0-\sin k_1)+i u}{2(\sin k_0-\sin k_1)-i u}.
\label{boson-S}
\end{equation}

\subsubsection{Fermionic walkers}
For the fermionic quantum walkers, we employ the nearest-neighbor
interaction \eqref{interaction}. The LHS of the 
Schr\"odinger equation \eqref{schroedinger2} at $x_0=x_1-1=:x$
gives
\begin{align}
\text{LHS}&=
\psi(x-1,x+1;k_0,k_1)+\psi(x,x+2;k_0,k_1)+
u \psi(x,x+1;k_0,k_1) \nn \\
&=\left[(e^{ik_1}+e^{-ik_0}+u)e^{ik_1}-S_{01}(k_0,k_1)
\left\{(e^{ik_0}+e^{-ik_1}+u)e^{ik_0}\right\}\right]e^{i(k_0+k_1)x}.
\end{align}
Also, we find the RHS of \eqref{schroedinger2} is given by
\begin{equation}
\text{RHS}=2(\cos k_0+\cos k_1)
\left\{e^{ik_1}-S_{01}(k_0,k_1)e^{i k_0}\right\}e^{i(k_0+k_1)x},
\end{equation}
and therefore 
\begin{equation}
S_{01}(k_0,k_1)=\frac{1+e^{i(k_0+k_1)}-e^{ik_1}u}{1+e^{i(k_0+k_1)}-e^{ik_0}u}.
\label{fermion-S}
\end{equation}

\end{appendix}

\bibliography{BibFile}

\end{document}